# Highly Flexible and Modular Simulation Framework for Magnetic Particle Imaging

P. Vogel[1*], M.A. Rückert[1], T. Kampf[1,2], V.C. Behr[1]

[1] Department of Experimental Physics 5 (Biophysics), University of Würzburg, 97074, Würzburg, Germany.
[2] Department of Diagnostic and Interventional Neuroradiology, University Hospital Würzburg, 97080, Würzburg, Germany.
[*] Corresponding author, email: Patrick.Vogel@physik.uni-wuerzburg.de

**Abstract:**
Simulations with high accuracy are an essential part of scientific research to accelerate the innovation process. They are especially useful for finding novel approaches or optimizing existing methods. Today, powerful software tools are available consisting of multiple packages with a wide variety of features, methods, and models meet the requirements for different questions in multiple fields of research. Unfortunately, the complexity and often inflexibility of such unspecific software tools can hinder an optimal workflow. Especially in the case of a novel research fields, e.g., Magnetic Particle Imaging (MPI), the requirements on a software tool are a high degree of flexibility paired with a manageable number of highly specific features to provide fast and easy access. Thus, often research sites generates their own software solution to address their specific demands.

Until now, only few simulation frameworks are available, which partially fulfil most requirements of the young field of MPI. However, the coverage of the entire process of emulating a full MPI experiment from magnetic field and particle dynamic simulations, hardware and sequence programming over signal generation and data processing to final reconstruction and visualization combined with an easy-to-use graphical user interface (GUI) and without the need of complex combination of different software packages cannot be found in the literature or MPI community at the time of this writing.

With the presented modular simulation framework consisting of multiple interconnected software packages to specific purposes, all necessary simulation steps are provided. The high degree of flexibility and modularity allow the simulation and emulation of almost any kind of MPI scanners known in the MPI community. Furthermore, the modular framework allows an easy connection of third-party software using dedicated interfaces between important steps.

## I. Introduction

Since the simulation and emulation as well as optimization of signals, hardware parts, processes and many more are an essential part in different fields of research and industry, a huge variety of software solutions for different research problems are available. In theory, almost any imaginable problem can be implemented, simulated, and emulated with the right software tool.

One major drawback is the complexity of such highly flexible software packages covering multiple parts of the entire process, e.g., from the simulation of novel hardware design of imaging devices to the final reconstructed image. This is a result of the different requirements in each processing step, that have to be handled individually.

For Magnetic Particle Imaging (MPI) [1], there are several simulation studies published, which use commercial software tools, but there are also specific software packages available [2-10] as well as frameworks for specific types of MPI scanners or open-source projects for educational purposes [11, 12]. Each software solution solves one or more specific questions in a small area. Unfortunately, these tools are often difficult adapt to the requirements at other research sites, which typically triggers the creation of another specific software tool. However, to overcome this issue, a modular framework offering a highly flexible workflow allowing the simulation and emulation of almost any question in the field of MPI is required. Open-source projects often provide open libraries consisting of a variety of functions and procedures to solve multiple problems, but the use of those libraries often requires additional frameworks or third-party software to execute which hampers its usability.

One solution is a software environment encapsulating all required libraries but with an easy-to-use graphical user interface. To ensure performance, the usage of a high-level programming language is mandatory, providing the highest flexibility but also the highest complexity to build a problem-specific software package.

## I.I. Brief introduction to MPI

Magnetic Particle Imaging (MPI) is a young tomographic technology [1], which has shown significant improvements in the last decade in hardware design and development as well as reconstruction methods [13]. MPI was successfully in different fields of research [14, 15] and makes it a promising candidate for future applications in medicine [16, 17].

MPI relies on the nonlinear magnetization response of magnetic nanoparticles (MNPs) to time-varying magnetic fields. For encoding, a strong gradient is generated and moved along specific trajectories through the volume of interest to determine the spatial distribution of the MNPs (tracer).

Several encoding schemes, such as field-free point (FFP) or field-free line (FFL) encoding, have been presented in the past and lead to multiple different designs and approaches for the generation of the required magnetic fields with more or less complex coil structures [13, 16, 18].

The signal acquisition in MPI devices is mostly done by inductive measurement of the MNP magnetization over time. The induction voltage $u$ within an electromagnetic receive coil (rx) with coil sensitivity $\boldsymbol{p}$ of a spatially distributed MNP ensemble with concentration $c$ inside a given field of view (FOV) $\Omega$ can be described by the MPI signal equation

$$u(t) = -\mu_0 \int_\Omega c(\boldsymbol{r}) \boldsymbol{p}(\boldsymbol{r})^T \frac{\partial}{\partial t} \overline{\boldsymbol{m}}(H(\boldsymbol{r},t)\boldsymbol{r},t) \mathrm{d}^3 \boldsymbol{r}, \qquad (1)$$

with $\mu_0$ as the vacuum permeability and $\overline{\boldsymbol{m}}$ as the averaged magnetic moment of the MNP ensemble. The magnetic moment strongly depends on the external magnetic field $\boldsymbol{H}(\boldsymbol{r},t)$, which makes it difficult to model this parameter appropriately [19]. To overcome this issue, the signal, or kernel of Equ. 1,

$$s(\boldsymbol{r},t) = -\mu_0 c(\boldsymbol{r}) \boldsymbol{p}(\boldsymbol{r})^T \frac{\partial}{\partial t} \overline{\boldsymbol{m}}(\boldsymbol{r},t) \qquad (2)$$

is measured or simulated [20, 21] at each position $\boldsymbol{r}$ in the FOV during a time-consuming calibration resulting in a discrete transformation matrix (system matrix) $\widehat{\boldsymbol{S}}$, which transforms the spatial distribution of the MNPs into the corresponding signal acquired with the very system the matrix is designed for. The linear system

$$u = \widehat{\boldsymbol{S}} \cdot c \qquad (3)$$

consists of the transfer function $\widehat{\boldsymbol{S}}$, the induction voltage $u$ and the desired particle distribution $c$ and can be solved by different approaches, e.g., iterative solvers or directly via matrix inversion to determine the spatial distribution of the MNP concentration $c$ [22].

The advantage of the system matrix approach is the high accuracy since all information, e.g., the specific field distortions of the scanner, the MNP characteristics and the receive chain dependent signal distortions, are considered and stored in the system matrix. On the other hand, it is highly inflexible, since even slight changes make a new calibration necessary [23].

Another reconstruction approach is the x-space reconstruction [23-29], which directly works with data in the time domain instead of the Fourier domain. For image reconstruction, the data are directly gridded at their corresponding spatial position on a 2D surface or in a 3D space resulting in a raw-image, where the MNP distribution is convolved by the system-specific point-spread function (PSF) [22, 27].

The advantage of x-space reconstruction is its high flexibility, simple implementation, and short calculation times, but slight distortions due to inaccurate magnetic fields can cause artifacts [23].

However, there are many methods and approaches to overcome these issues, e.g., model-based system matrix, image-based system matrix, hybrid reconstruction approaches, single harmonic methods and many more [21, 30-34].

## I.II. Requirements for MPI simulations

The entire process for an adequate simulation of, e.g., a complete MPI device with the goal to emulate the reality as well as possible, requires multiple sophisticated steps. In the past, many software packages have been presented fulfilling specific requirements and offering appropriate simulation and/or emulation results for the specific application [2-12]. However, often the flexibility is limited, and the handling is complex due to the absence of a graphical user interface (GUI). For a complete emulation of an MPI system, multiple features should be provided by the framework.

The *basic workflow* is listed in the following:
- **Magnetic Field Calculator**
  Based on defined coil structures, the magnetic field is calculated for selected points in space. Since the frequency range is below 10 MHz a Biot-Savart calculator [35] can be used. This simplifies the implementation and reduces the computational effort – even more for multiple points and long sequences (parallelization).
- **Magnetic Nanoparticle (Magnetization) Models**
  For the generation of a realistic behavior of MNP magnetization in (time-varying) magnetic fields, several models are used: From simple single particle model (SPM, Langevin-function) over more realistic models with stochastic terms (Langevin equation) to PPI (particle-particle-interaction) approaches. Depending on the topic and available calculation times highly sophisticated and complex MNP signals can be generated [36-40].
- **Data Acquisition**
  The change in magnetization is measured inductively in real MPI devices, thus the framework offers an induction calculator for custom receive coils following the reciprocity theorem for arbitrary geometries known from NMR [41, 42] (see Sec. II.I.III.).
- **Data Processing**
  Real signals are affected by noise and distortions in amplitude and phase, e.g., due to hardware limitations. Therefore, digital filtering and data processing is essential for signal cleaning and preparation for further steps [7].
- **Reconstruction**
  After data preparation, several reconstruction methods (gridding, deconvolution, system matrix, etc.) can be applied to visualize the data [23].
- **Visualization**
  Appropriate data visualization is essential for analyzing. This often requires sophisticated tools, e.g., for 3D volume rendering [8, 24].

In addition to the listed basic features a simulation framework for MPI should provide, user experience will be improved by
- **Operation and Handling**
  A fast and intuitive workflow can be essential for innovative solutions and/or optimizing common concepts. An easy-to-use 3D-GUI as well as near real-time visualization and data preview are essential features for that.
- **Adaption and Flexibility**
  Closed all-in-one frameworks often provide good press-button solutions. In novel fields it is more important to replace parts of a data processing pipeline for a more optimized workflow. For that, a modular framework consisting of several packages can be advantageous. Standardized and open interfaces should be available to transfer data between the packages.
  For higher flexibility, a proprietary scripting language allows the direct access to functions for data processing, signal generation, etc. and allows automatization of complex processes (see Sec. II.I.VII. Scripting tool).
  With cross-platform capability, the software packages are available on multiple platforms and operating systems.
- **Response, Speed and Maintenance**
  For optimal performance, a high-level programming language permanently maintained over long times and providing embedded compilers for multiple operating systems is required, e.g., Rad Studio (Embarcadero, USA). Using optimized libraries, e.g., FFTW [43] for fast Fourier transformation, LABLAS or LAPACK [44] for fast matrix operations, the fastest possible data processing can be provided for maximum responsiveness of the GUI. In addition, parallelized implementation of parts of the source code and pre-calculation steps (see Sec. II.I.VII. pre-calculation) can dramatically increase performance.

**II. Material and methods**

The proposed MPI simulation framework consists of three different software packages (see Fig. 1), which can be used individually or together via specific interfaces, e.g., memory mapped files (MMF):
- **Magnetic Field Simulator (MFS)**
  Main simulation environment for the generation of MNP signals of arbitrary magnetic field configurations (Fig. 1a).
- **Reconstruction Framework (RiFe)**
  Framework optimized for fast data processing and near real-time image reconstruction (Fig. 1b) [7].
- **3D Visualization Tool (3DVT)**
  Software package optimized for visualization of 3D data using volumetric rendering (Fig. 1c).

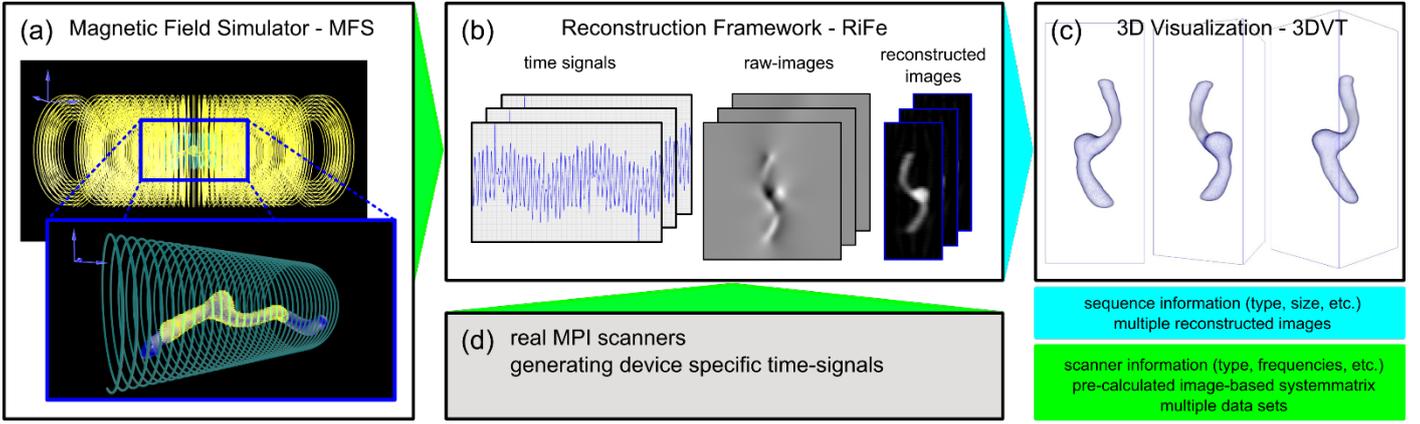

**Fig. 1:** *Sketch of the framework and their connections shown on an example of a 3D simulation of an aneurysm phantom scanned with a TWMPI scanner: (a) In the **MFS**, the entire scanner is simulated, and the data sets are generated. (b) Each data set is processed and reconstructed (**RiFe**) before visualization with the 3D visualization tool (**3DVT**) (c). (d) Instead of processing the input from the simulation framework (MFS), the time signals can be provided directly from a real MPI scanner. Green/cyan: Between the modular software packages, specific information can be transferred via specific interfaces, e.g., MMF.*

Each software module is programmed in Delphi® (Embarcadero RAD Studio 11 Alexandria, USA), a general-purpose programming language providing an integrated development environment (IDE) for rapid application development on multiple platforms and operating systems.

For data transfer between the software modules, either specific data formats (time data, Fourier data, image data, system-matrix data, etc.) can be used or a fast data exchange through Memory Mapped File technology (MMF) is available.

All modules are user-friendly designed (3D graphical user interface, etc.) and optimized for fast data processing (Magnetic Field Simulator) and for near real-time visualization with low latency (Reconstruction Framework and 3D Visualization Tool). The latter feature in combination with specific data transfer protocols (e.g., Teledyne LeCroy, USA or Tektronix, USA) allows the near real-time reconstruction and visualization of data streams obtained on real MPI scanners (Fig. 1d) [7, 24, 45, 46].

**II.I. Magnetic Field Simulator – MFS**

The MFS tool is a versatile and flexible simulation environment for magnetic fields in 4D [5]. A 3D-GUI with intuitive controlling (mouse and/or multi-touch) provides a fast and easy navigation around the designed structures. Standard features known from versatile 3D software are available, e.g., specific lighting or different rendering settings.

The MFS software is structured in different types of helper-containers with different assignments. This allows a standardization and thus optimization of data processing.

**II.I.I. Conductor-container**

For the simulation of quasi-static and dynamic magnetic fields with frequencies below 10 MHz, the law of Biot-Savart is valid, which is sufficient for common MPI experiments [35]

$$\mathrm{d}\boldsymbol{B}(\boldsymbol{r}) = \frac{\mu_0}{4\pi} I \mathrm{d}\boldsymbol{l} \times \frac{\boldsymbol{r}-\boldsymbol{r}'}{|\boldsymbol{r}-\boldsymbol{r}'|^3}, \tag{4}$$

with $\mathrm{d}\boldsymbol{B}(\boldsymbol{r})$ being the magnetic field at position $\boldsymbol{r}$ in 3D space generated by a flexible current $I$ due to a differential element $\mathrm{d}\boldsymbol{l}$ at position $\boldsymbol{r}'$ (see Fig. 2 top left) and $\mu_0$ as the vacuum permeability.

Based on this differential concept, arbitrary coil designs for magnetic field generation can be generated element by element (see Fig. 2). Parameterized function-based templates, e.g., from simple single loops over saddle-coil pairs of canted-cosine-theta (CCT) coils to toroidal structures, or sophisticated structured assemblies, e.g., as spherical arrangement are available [47]. Each conductor-container consists of an array of size $N_{\mathrm{el}}$ with all element information and a transformation matrix $\widehat{M}$ (see Sec. VI.S1)

$$\mathrm{conductor}(\widehat{M}, N_{\mathrm{el}}) = \{\widehat{M}, \{\{\boldsymbol{r}, \boldsymbol{dl}\}_1, \{\boldsymbol{r}, \boldsymbol{dl}\}_2, \ldots \{\boldsymbol{r}, \boldsymbol{dl}\}_{N_{\mathrm{el}}}\}\}. \tag{5}$$

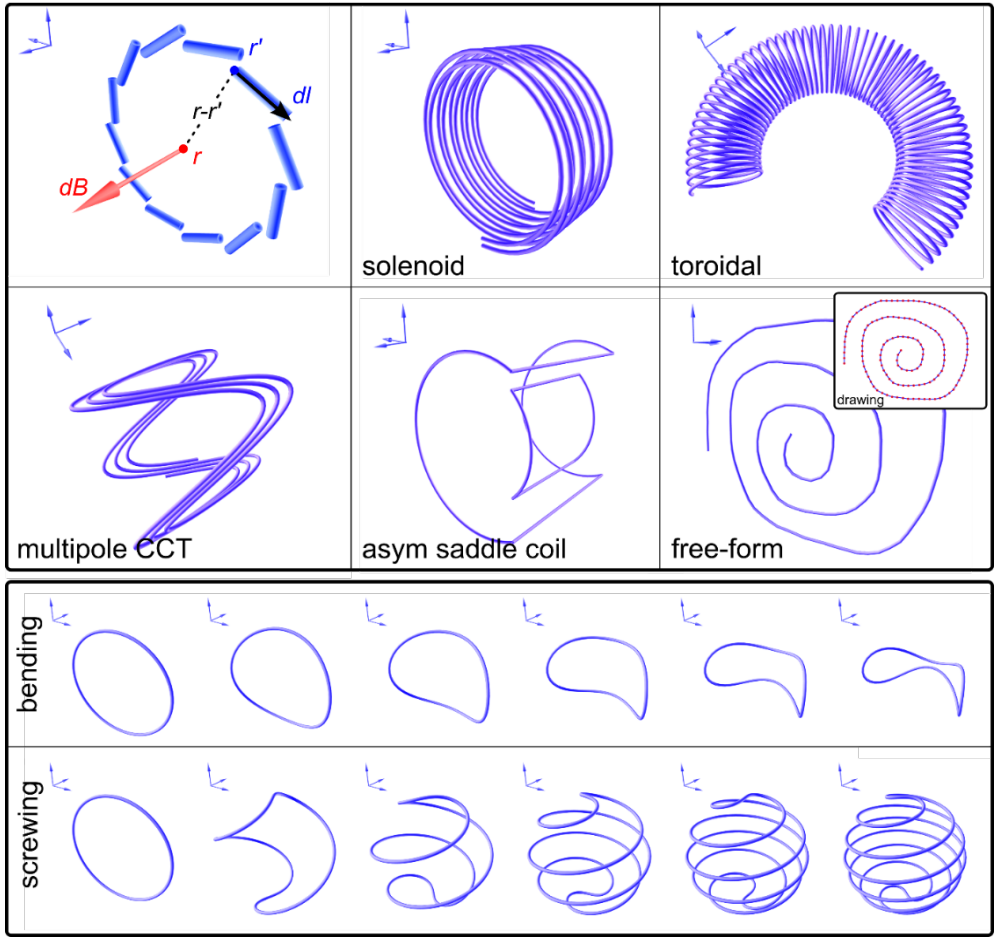

*Fig. 2: Top left: magnetic field calculation using Biot-Savart law: for each element **dl** within a conductor array, the partial magnetic field **dB** at a specific position **r** can be calculated. Top: With this differential concept, arbitrary structures can be generated element-wise, e.g., solenoidal, toroidal, multipole CCT, saddle coils or free-form shapes. Bottom: Additional features such as bending or twisted offers the generation of complex designs.*

Standard features, such as translation, rotation and scaling, but also special functions, such as bending or twisting, can be applied to the conductor elements to modify position and shape and thus the resulting magnetic field (Fig. 2 bottom). In addition, for more individual structures, a tool is available for hand-free drawing of conductor structures. This can be helpful for rapid prototyping structures and/or also for educational purposes. By combining multiple conductors to one single element (conductor network list – cnl-element), complex structures can also be stored and used in other projects.

Each conductor can be used in two basic modes: as **transmit coil (tx)** or as **receive coil (rx)**. In **tx mode**, the conductor exclusively serves as magnetic field generator and generates a magnetic field following Equ. 4 & 6 at selected positions with a given arbitrary time-dependent current sequence $I_m(t)$. At each position $r_i$, the magnetic field $B(r_i, t)$ generated by all transmit conductors $N_{\text{cond}}$ each consisting of $N_{\text{el},m}$ single elements is calculated for every time step $t$ (see Sec. II.VII.)

$$\boldsymbol{B}(\boldsymbol{r_i}, t) = \frac{\mu_0}{4\pi} \sum_{m=1}^{N_{\text{cond}}} \sum_{n=1}^{N_{\text{el},m}} I_m(t) \mathrm{d}\boldsymbol{l_{m,n}} \times \frac{\boldsymbol{r_i} - \boldsymbol{r_{m,n}}}{|\boldsymbol{r_i} - \boldsymbol{r_{m,n}}|^3} \quad . \tag{6}$$

In **rx-mode**, this conductor is connected over a magnetic nanoparticle array (see Sec. II.I.IV.) to a magnetic field array (see Sec. II.I.IV.) collecting magnetic field and particle information (magnetization) generated by the transmit conductors and particle system at these points.

The dependency graph in Fig. 3 shows a simple MFA, a slice-container and an rx coil. The transmit coils are used for magnetic field calculation at the specific points represented by the MFA-container. According to their function, i.e., slice-container or receive conductor, further data processing is performed either for visualization of the magnetic fields or calculating the induction signal for a given receive coil and a given MNP ensemble.

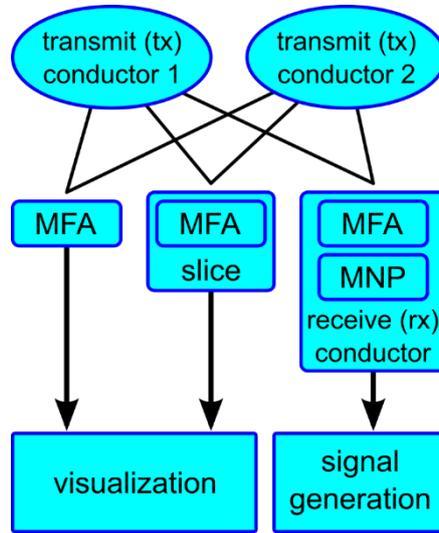

*Fig. 3: The MFA-container is the central part of the simulation. Depending on their function, e.g., MFA visualization, slice visualization or signal generation for a receive coil, further processing steps are performed.*

**II.I.II. MFA-container**

MFA stands for ***magnetic field array*** and represents an ensemble of $N$ dedicated points $r'$ in 3D space. The MFA is a container, which can be used as a modular computation unit of all parameters required for calculation and visualization of magnetic fields from different sources. For each position $r'$ several additional information is stored, such as the latest magnetic field vector $B$, the previous magnetic field vector $B'$, a pick-up table $\hat{C}$ consisting of information about associated conductors

$$\text{MFA}[N] = \{\{r',B,B',\hat{C}\}_1, \{r',B,B',\hat{C}\}_2, \dots, \{r',B,B',\hat{C}\}_N\}. \quad (7)$$

The MFA-container has an optimized data structure for fast data processing and calculation, as well as interfaces for data visualization. E.g., several parts of the entire process are parallelized and can be pre-calculated in the background to ensure rapid data generation and highly responsive interaction experience of the user (see Sec. II.I.VII.).

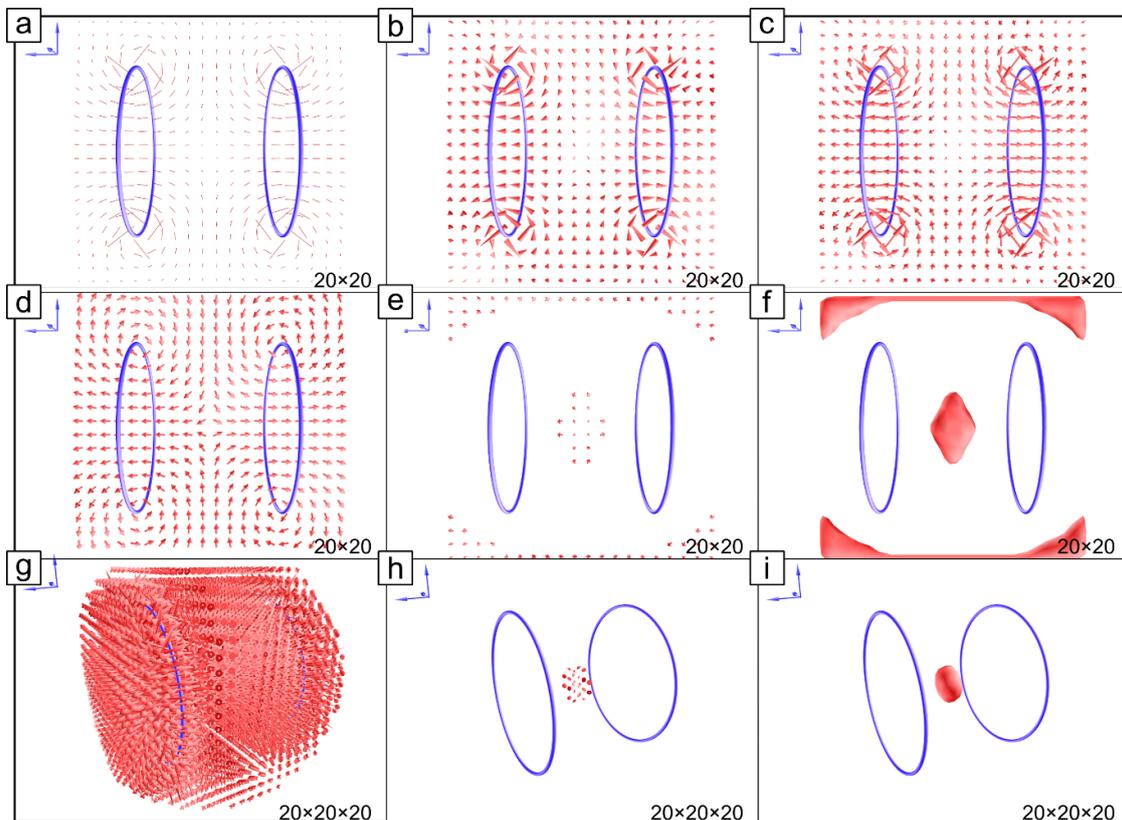

*Fig. 4: Different methods for MFA visualization: **(a)-(c)** several types of magnetic field vector length rendering, displayed as unit vectors **(d)** or with specific thresholding **(e)** in 2D or 3D **(g) & (h)**. Using marching cube algorithm, iso-surfacing can be used for detailed visualization of magnetic field geometries.*

For visualization of MFAs, the framework offers different methods as indicated in Fig. 4. From simple magnetic field vector renderings (Fig. 3 a to d) to solid renderings (iso-surfacing) with specific constraints, e.g., thresholding or unit vectors, in 2D or 3D (Fig. 4 e to i). In addition to magnetic field vector information, the visualization of magnetic field gradients ($dB/dr$ or $dB/dt$) or special information about the magnetic field behavior over time, e.g., circular, elliptic or linear magnetic fields, is available.

**II.I.III. Slice-container**

A specific type of the MFA-type is the slice-container, which is a 2D plane with a given size (width and height). It can be positioned and rotated arbitrarily in the 3D space to visualize, e.g., the magnetic field or gradient with a given number of voxels $N_x$ and $N_y$ spanned by the plane vectors $u$ and $v$. The slice-container consists of an MFA-container and additional parameters, e.g., affine transformation matrix $\widehat{M}$ (see Sec. VI.S1) and more ($PAR$)

$$\text{slice}(\widehat{M}, N_x, N_y) = \{\widehat{M}, u, v, MFA[N_x \cdot N_y], PAR\}. \tag{8}$$

In Fig. 5 several different visualization methods are shown indicating various interesting parameters of static and dynamic magnetic fields generated a given conductor array. Fig. 5b&c shows an example of two perpendicular solenoids driven by the same sinusoidal current but with a phase shift of 90° generating a rotating magnetic field. While Fig. 5b shows the absolute values of the magnetic field vector, Fig. 5c shows: the magnetic field gradient $dB/dr$ (top left), the magnetic field derivative $dB/dt$ (top middle), absolute values of phase (top right), absolute phase values along x-direction (bottom left), specific contour plot (bottom middle) and special magnetic field parameters, e.g., circular behavior over time (bottom right). Fig. 5d-e shows the absolute magnetic field of a static spiral-shaped conductor. With profile plots more specific information can be extracted (Fig. 5f). In Fig. 5g-i a contour plot visualization with increasing slice resolution shows the increasing image quality at the expense of prolonged calculation times.

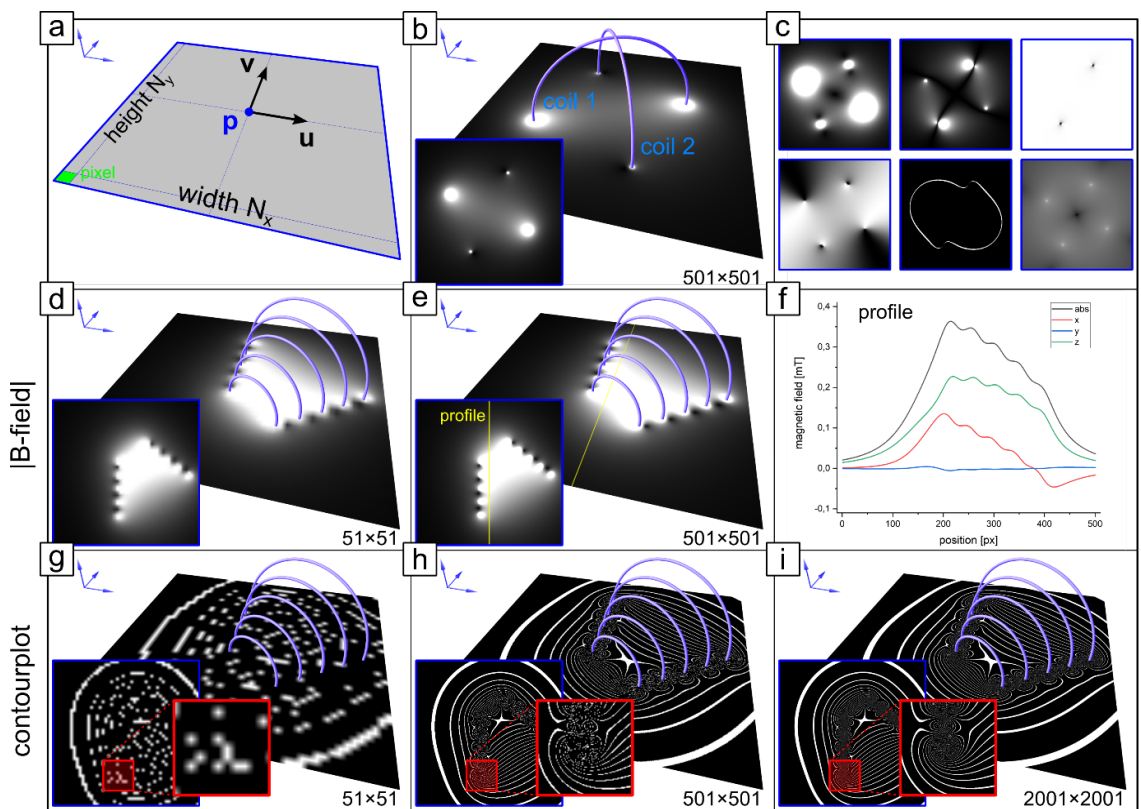

***Fig. 5:*** *(a) Sketch of a slice-container with specific size (width and height) positioned in 3D space with $p$ as support vector and $u$ and $v$ as span vectors. (b) The absolute magnetic field values at a point in time of two perpendicular solenoids running with a phase shift of 90° generating a rotating magnetic field. (c) Visualization of different parameters: top left: magnetic field gradient $dB/dr$, top middle: magnetic field derivative $dB/dt$, top right: absolute phase values, bottom left: absolute phase values along x-direction, bottom middle: specific contour plot, bottom right: magnetic field circulation behavior over time. Absolute magnetic field values of a spiral-shaped conductor with different slice-resolutions: 51×51 px² (d) and 501×501 px² (e). (f) 2D profile plots along the center shows more information about the magnetic field. (g)-(i) Contour plots for different slice sizes (51×51 px², 501×501 px², 2001×2001 px²).*

## II.I.IV. MNP-container

For MPI simulations, a mandatory feature is the particle ensembles (MNPs) and their behavior as well as their spatial distribution. Like the slice-container, the MNP-container consists of an ensemble of multiple discrete points in space representing the positions of particle ensembles ($\widehat{M}$, MFA[$N$]) following a specific magnetization model $\widehat{m}$

$$\text{MNP}(\widehat{\boldsymbol{M}}, N) = \{\widehat{\boldsymbol{M}}, \widehat{\boldsymbol{m}}, \text{MFA}[N], \text{PAR}\}. \tag{9}$$

For calculating the nonlinear behavior of the magnetization $\bar{m}$ of MNP ensembles to a time-varying magnetic field $H(r,t)$ multiple models are available [37-40] The mostly used model in the MPI community is the Langevin function $L(\xi)$ (Eqn. 10) with $\mu_0$ as the vacuum permeability, $k_B$ as the Boltzmann constant and $T$ as temperature. With the Langevin parameter $\xi$ this model describes paramagnets in external magnetic fields semi-classically [37]

$$L(\xi) = \coth(\xi) - \frac{1}{\xi} \ \text{ with } \xi = \frac{\mu_0 m H}{k_B T}. \tag{10}$$

The entire magnetization $\bar{m}$ of the ensemble can be determined by $\bar{m} = NmL(\xi)$ with $N$ as the number of particles within the volume elements of the simulation. This model follows the single particle model (SPM), which assumes, that the magnetic moments of the particles follow the external magnetic field instantaneously [37, 48]. This makes it easy to implement and quite fast.

The second model is a naive implementation of MNP relaxation using a lag-effect (Eqn 11). For that, the magnetization of the MNPs is interpolated from the current magnetization vector and the last *N* magnetization vectors. This results in a relaxation type, which requires less computational effort as the third one

$$\boldsymbol{m}_t = \frac{1}{N+1} \sum_{i=0}^{N} \boldsymbol{m}_{t-i}, \tag{11}$$

A third model follows the Langevin equation (Eqn. 12), which overcomes limitations of the Langevin function approximation, since it considers the dependency of the magnetic response on viscosity of the suspending liquid and changes in the particle's hydrodynamic diameter [49, 50]

$$\frac{d\boldsymbol{m}}{dt} = \frac{1}{\zeta}(\boldsymbol{m} \times \boldsymbol{H}) \times \boldsymbol{m} + \sqrt{\frac{2k_B T}{\zeta}} \boldsymbol{\lambda} \times \boldsymbol{m}, \tag{12}$$

with $\boldsymbol{\lambda}$ as normally distributed random vectors with expectation value $\langle \lambda_i(t) \rangle = 0$ and $\zeta = \kappa \eta R^3$ as the Stokes-Einstein diffusion coefficient (viscosity-coefficient) which depends on the viscosity $\eta$ of the surrounding liquid, the particle diameter $R$ and a shape factor $\kappa$. The Langevin equation consists of a deterministic part and a stochastic part. This means that for the calculation of the magnetization multiple single magnetic moments have to be tracked over time. This results in a more complex implementation and requires more computational effort. This model is also inly valid for particles that show no Néel relaxation, i.e., the magnetization is fixed inside the particle.

In Fig. 6 the magnetization for three different models is shown in a simple experiment with parallel transmit and receive coils. A rectangular shaped current pulse generates a strong magnetic field. The Langevin function instantaneously follows the shape of the provided current. A naive relaxation model, which simply interpolates the last $N$ magnetization vectors, shows a symmetric relaxation behavior [50]. The Langevin equation model, here calculated from $50k$ single particles, shows a more realistic relaxation. The calculation times for the Langevin function model and the Langevin equation model differ by four orders of magnitude.

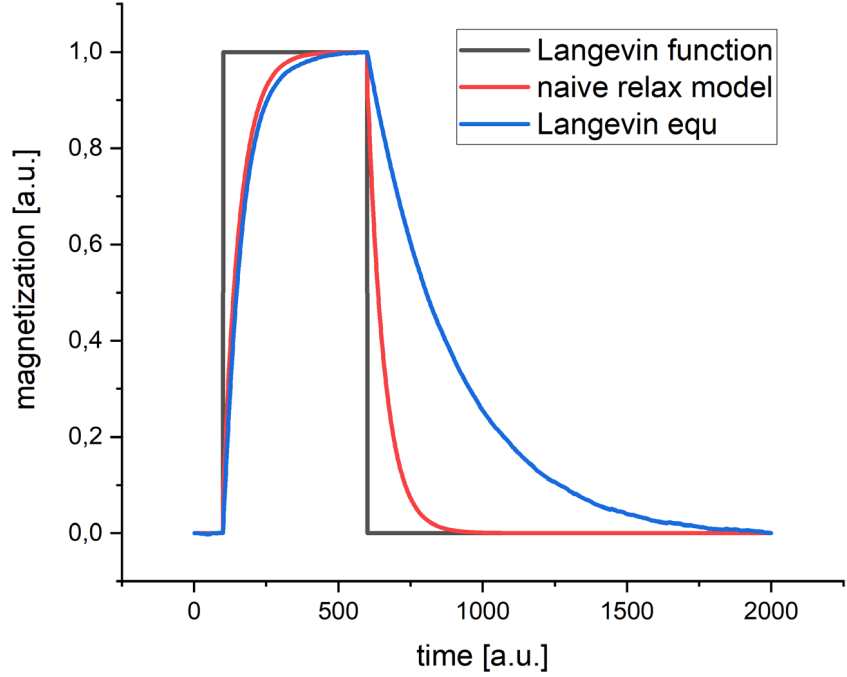

**Fig. 6:** *Plot of different magnetization models demonstrating the behavior of MNPs during a rectangular shaped magnetic field pulse. The Langevin function follows instantaneously the external magnetic field, the naïve relaxation model shows a symmetric magnetization behavior for induced magnetization and free relaxation and the magnetization following the Langevin equation shows the most realistic behavior [50].*

*-Alternative magnetization calculator – dynamic Bloch solver*
Furthermore, the magnetization response of a spin ensemble or isochromate in dynamic magnetic fields can be calculated using a dynamic Bloch-solver [51].

$$\frac{d\boldsymbol{m}}{dt} = \gamma \boldsymbol{m} \times \boldsymbol{H} - \frac{\boldsymbol{m}_\perp}{T_2} - \frac{\boldsymbol{m}_\parallel}{T_1}$$
$$with\ \boldsymbol{m} = \boldsymbol{m}_\perp + \boldsymbol{m}_\parallel\ ,\quad \boldsymbol{m}_\parallel \parallel \boldsymbol{H}\ ,\quad \boldsymbol{m}_\perp \perp \boldsymbol{H} \tag{13}$$

This allows not only the simulation of NMR (nuclear magnetic resonance) or MRI (magnetic resonance imaging) signals with static $B_0$ fields, but also the magnetization response on arbitrarily magnetic field dynamics.

**II.I.V. Receive coil-container**
The receive coil-container is a sub-structure of the conductor-container consisting of an additional MNP-container (see Fig. 3)

$$rx(\widehat{\boldsymbol{M}}, N_{el}, N) = \{conductor(\widehat{\boldsymbol{M}}, N_{el}), MNP(\widehat{\boldsymbol{M}}, N), \{\boldsymbol{p}_1, \boldsymbol{p}_2, \dots, \boldsymbol{p}_N\}\}. \tag{14}$$

This MNP-container, or better the dependent MFA-container serves as array, which defines the position of sensitivity map or points $\boldsymbol{p}$ (cf. Eqn. 1) of the selected coil. Following the NMR reciprocity theorem, which state that the magnetic field generated by a coil at specific points in space are equivalent to its sensitivity in the same spot [41, 42], the calculation of the induced voltage $u(t)$ in this receive coil can be performed by summing up the magnetization changes over time $\frac{\partial}{\partial t}\overline{\boldsymbol{m}}(\boldsymbol{r}, t)$ in every point generated by a transmit coil (Faraday's law).

Fig. 7 shows the projection calculation for a single point in space defined by the MNP-container. The magnetic field $\boldsymbol{B}$ generated by the transmit coil is used to calculate the magnetization $\boldsymbol{m}$ at this point following the selected magnetization model (MNP-container). The induced voltage $u$ is the change of the projected magnetization $\boldsymbol{m}$ on the sensitivity map $\boldsymbol{p}$.

Instead of the induced voltage, also the magnetization can be directly extracted as signal as well as the vectorial signal of the magnetic field generated of the transmit coils.

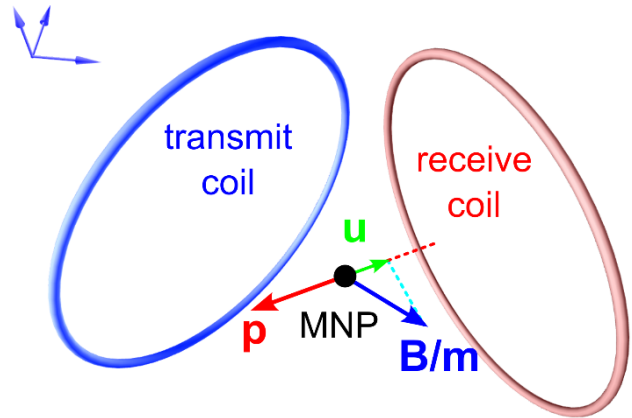

*Fig. 7: Projection calculation to get the induced voltage u or signal within a receive coil: the generated magnetic field **B** of a transmit coil at a given point in space defined by an MNP array is projected onto the sensitivity map **p** of a receive coil.*

**II.I.VI. 3D model-container**

This container is a passive container used for visualization of 3D models as well as for calculation of MNP distributions based on 3D volumes. The container can import stl files (standard triangle language), which is a standard file format of various 3D software, e.g., for rapid prototyping, 3D printing and computer-aided manufacturing. It uses lists of triangles to describe the surface of the 3D model, which can be rendered as triangle mesh (see Fig. 8 a).

For calculation of 3D points, that are located inside the volume of the 3D model, a slicing algorithm (slicer) is used, which is well known in 3D printing. Based on a given start orientation multiple parallel 2D slices are generated at a given spacing. For each slice, which serves as 2D gridding field with given distances, the intersection with the 3D model is calculated in form of a closed 2D contour. Using a point-in-polygon algorithm, all 2D points on this slice lying inside the contour can be determined as indicated in Fig. 8 b. These 3D points can be transferred to an MNP-container, where the sample emulation can be used as MNP distribution with specific magnetization models.

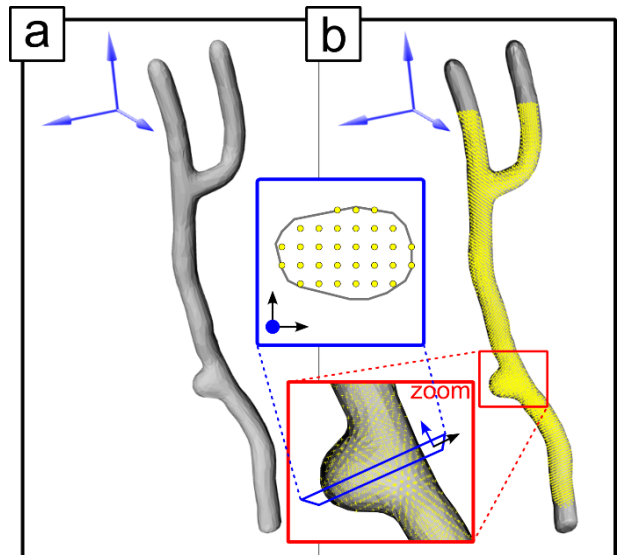

*Fig. 8: (a) 3D rendering of a 3D model using triangulation. (b) Utilizing 3D printing slicer algorithms, discrete 3D points can be calculated within the 3D model to emulate a magnetization distribution (MNP-container). By slice-wise cutting of the model within a closed 2D contour, the MNP positions can be calculated.*

**II.I.VII. Calculation process**

An important feature of any simulation software with a GUI is the response time, which means the update time, between consecutive simulation time steps to ensure a fluid presentation and visualization of dynamics and results, e.g., time-varying magnetic fields. Furthermore, an optimized calculation process is important for fast generation of huge data sets, which can be used for further processing.

In the MFS software, several calculation methods are available, which separate independent steps allowing to update them on demand only or combine similar steps to optimize calculation time through parallelization.

For each container type an individual update-function is implemented, which allows to perform necessary updates selectively:

- *Update conductor-container (conductorlist)*
  This procedure updates on demand all parameters of each conductor, such as position, rotation, scaling (affine transformation matrix values), shape, type, etc. Depending on this, a pre-calculation process is started to calculate values and parameter, which have not necessarily to be updated during each time step, such as affine transformation matrix values, currents, etc. and/or dependent MNP/MFA-containers (see Sec. II.I.VII.). Additional pre-calculations are performed for generating the triangle mesh for 3D visualization.
- *Update receive coil-container (receive-coils)*
  A special sub-calculation process is the calculation of receive coils. If a conductor-container is defined as a receive coil, an additional pre-calculation step is performed: a sensitivity map (sens-maps) is created for each receive coil based on the dependent MNP-container, which is mandatory for fast calculation of signals over long time (data sets).
- *Update MNP-container (mnplist)*
  Depending on the selected magnetization type additional pre-calculations to the affine transformation matrix values of connected MFA-container are mandatory, such as initialization of random vectors.
  Additional pre-calculations are performed for 3D visualization.
- *Update slice-container (slicelist)*
  Updating the slice-container means triggering updating the connected MFA-container with new vector positions and directions required for 3D visualization.
- *Update MFA-container (mfalist)*
  The main function of the MFA-container is the management of the magnetic field vectors at dedicated positions required for MNP-container, slice-container, and receive coil-container. Thus, the update is often triggered by these containers. In addition, the MFA-container is used for 3D visualization of magnetic field parameters, which requires among other the pre-calculation of triangle-mesh models for visualization of 3D arrows.

*-Generation of data sets*
This special procedure works independently of the above-mentioned calculation-for-visualization process. For the generation of data sets consisting of multiple time steps, an encapsulated data processing step has been implemented, which provides several methods for time step calculation, especially for parallelization:
1. The first parallelization method is used for the generation of a single data set with multiple MNP-containers connected with MFA-containers. For each individual MFA-entry (3D point) a separate thread is generated calculating the full data set coming from this specific point in space. At the end, all single MNP data sets are summed up.
2. The second parallelization method is similar to the first one and used for generating multiple data sets at once. This is useful for the calculation of system matrices, which can consist of multiple positions in space where a data set has been calculated for a point-like sample. The created data sets can be stored separately or further processed to create the final system matrix.
3. The third parallelization method is used for time-dependent calculation processes like flow simulations or particle-particle-interaction (PPI) simulations, and tries to parallelize for each time point all required data of conductors, MNPs, MFAs, etc.
4. A fourth parallelization method has been implemented within the calculation step of Langevin-equation. Since for each time step multiple single magnetization vectors $N_{stoch}$ are used for calculating the stochastic part of the signal, it can be useful to parallelize these calculations by dividing all vectors in $M$ threads consisting of $N_{stoch}/M$ vectors each. This method can be combined with any other method.

There are additional parts of the source code, where parallelization can be implemented, e.g., the pre-calculation steps. All threads are monitored and managed by a thread-pool, which is specifically set up for the dedicated hardware, the software is running on.
The generated data sets can be further processed within the MFS software (see Sec. II.I.VII. scripting tool), directly stored in a proprietary data format (*.mpi & *.dbl – see Sec. VI.S2) or provided to other software packages, e.g., to RiFe (Sec. II.II.), via fast MMF interface.

*-Pre-calculation*
Imagine a conductor structure with $N_{cond}$ conductors with $N_{el}$ elements each, which is positioned and oriented arbitrarily in space represented by matrix $\widehat{M}$. For calculation of the magnetic field $B(r_i, t)$ at $N$ specific points in space $r_i$ and time $t$ following Equ. 6, the position $r_{m,n}$ as well as the direction $dl_{m,n}$ of each single element must be calculated. A naive estimation results in the requirement of $4 \cdot 4 = 16$ multiplications and $4 \cdot 3 = 12$ additions per vector. In total, $N \cdot N_{cond} \cdot N_{el,m} \cdot 2 \cdot 16$ multiplications and $N \cdot N_{cond} \cdot N_{el,m} \cdot 2 \cdot 12$ additions are required. Since each mathematical operation requires computational time for calculation (addition requires typically 2 cycles and multiplication 7 cycles [52]), $N \cdot N_{\text{cond}} \cdot N_{\text{el,m}} \cdot 272$ cycles are required per time point. This does not sound like much in times of GHz processor, but for large datasets, e.g., with length $2 \cdot 10^6$, a 3D volume consisting of $10 \times 10 \times 10$ points in space and 10 conductors with 100 single elements each, at the end $272 \cdot 10^{12}$ cycles are required.

Often, there is no need to perform all calculations at each time step. E.g., when using spatially fixed conductors, the matrix calculations mentioned above only have be performed once and can be used for all time steps. This pre-calculation step checks for several processes whether a new calculation is required or not. If so, then the data are calculated and stored in separate pickup lists, which are updated on demand, to be used in further calculations.

The mentioned example with pre-calculating the 3D points for different tasks requires the major part in computation time in the software and is used in different situations, e.g., position and direction vector calculations, determining the sensitivity map or preparing a current look-up table over a dataset with given data length.

Further time-saving pre-calculations can be made for calculations of Fourier transforms by pre-calculating plans for specific data length [43].

### -scripting tool

A built-in scripting tool allows to run customizable scripts for automatization within the software, which can be used for optimization, visualization, and demonstration of magnetic fields and MPI or MRI signals. All parameters of the containers, e.g., changing the positioning or particle properties, can be manipulated within the scripting area. Furthermore, signal processing commands, such as high-, low- or bandpass filtering, Fourier or Hilbert transformations, mathematical operations, or reconstruction commands, such as gridding processes are available to process data sets. The scripts can be stored in a proprietary format (see Sec. VI.S2).

### II.I.VIII. Useful MPI specific features

As mentioned at the beginning, there are multiple software frameworks available, which can be used for calculation and visualization of magnetic fields. However, especially specific questions, e.g., calculating system matrices or calculating the magnetization response of dedicated magnetic materials within a given receive coil, often are difficult to implement. Specific modes implemented within the MFS software allow the visualization of the magnetic fields and gradients and provides a highly responsive environment to investigate different feature, such as FFP or FFL surface rendering and 3D tracking, etc.

In the following, a selection of MFS features especially designed for MPI is shown:

- ***Calculation of non-linear magnetization response***

    The basic concept of MPI is the non-linear magnetization response of magnetic material to time varying magnetic fields [1]. Thus, a basic feature is an intrinsic and fast calculation for signals of that kind. In Fig. 9, time signals and hysteresis curves for different kind of MNP magnetization types (see Sec. II.I.IV) are shown, from pure Langevin function particles without Hysteresis to signal simulations with relaxation times (Langevin equation).

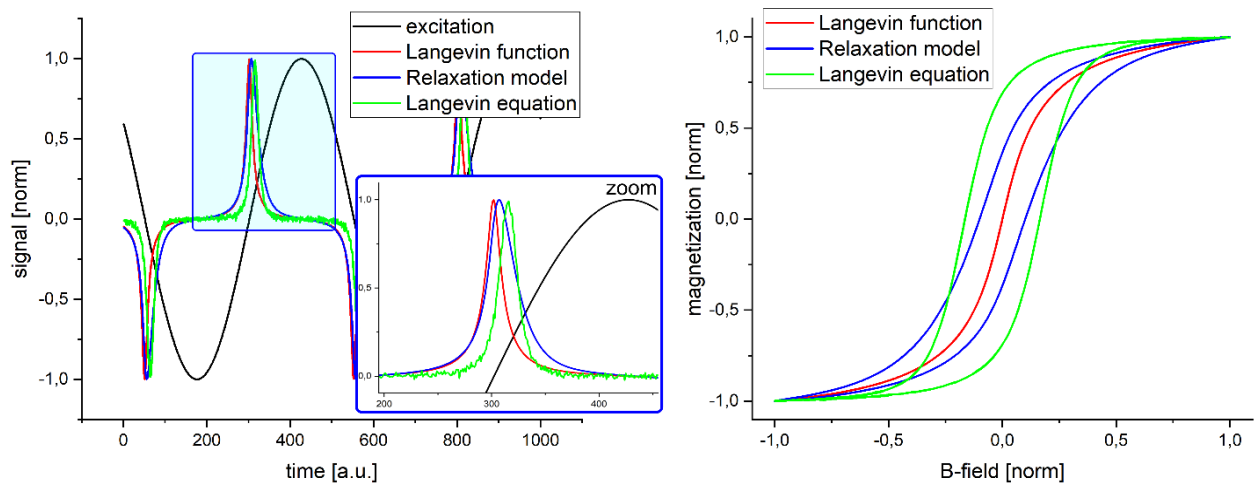

*Fig. 9: **Left:** Time signals showing the derivative of non-linear magnetization responses for different magnetization algorithms. **Right:** the corresponding hysteresis curves.*

- ***Calculation of system matrices***

  System matrices can be used for data reconstruction in MPI. Especially for real MPI scanners, a point-by-point acquisition of the corresponding system matrix is useful for black-box reconstruction. Since the acquisition time of real system matrices with high spatial resolution using a robot is quite long [1], hybrid system matrices using appropriate simulation models as well as pure model-based system matrices [21] are of high interest. With a highly optimized data processing, e.g., parallelization, the calculation time for huge system matrices can be dramatically reduced allowing fully automatic parameter optimization with short feedback times.

  For fast calculation within the MFS software, there is a specific function implemented providing calculation of time signals, Fourier transformed signals as well as dedicated system matrices: the user defines the number of system matrix points by setting up an MNP-container with, e.g., 21×15×1 for a rectangular 2D slice or 21×21×21 for a cubic 3D volume. For each MNP entry, a separate data set is calculated using an optimized parallelization process.

  In Fig. 10, an example is shown for generating a 2D system matrix for real-time visualization with RiFe. First, the size of the desired system matrix area is set (MNP-container – 25×25). Afterwards, the system matrix (ARSM – arbitrary system matrix) is generated with the specific parameters the MPI scanner is operating. Using an interactive script (see Sec. II.I.VII. scripting tool), the system function can be visualized and investigated before using the system matrix for image reconstruction (see supplementary video 2: system matrix).

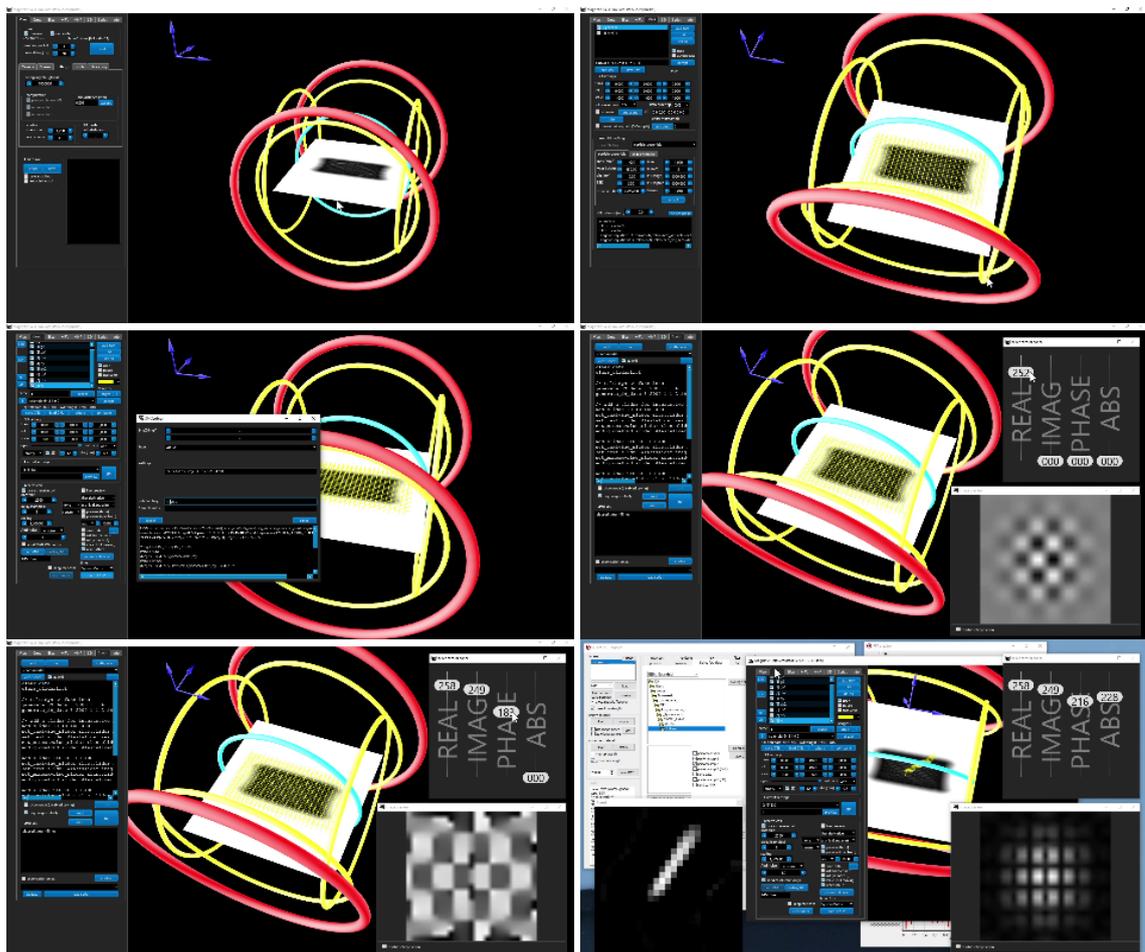

*Fig. 10: Process of generating a system matrix for virtual Gleich/Weizenecker-MPI scanner [1]: Top left: the area the FFP can cover is visualized with a slice-container and is used for setting up the system matrix size using an MNP-container with 25×25 entries and a spatial resolution of 0.50×0.25 mm² (top middle). Top right: setting up the system matrix type with dedicated information about the scanners frequencies (here $f_1$=10 kHz and $f_2$=11 kHz). Using an interactive script, the system function can be investigated (bottom left & middle) before image reconstruction within the RiFe software using the generated system matrix (bottom right).*

- ***Visualization of magnetic field parameters***

  For understanding of given scanner approaches or for investigating novel encoding schemes, a highly flexible visualization portfolio of magnetic fields and magnetic field gradients are important (see Sec. II.I.III. & Sec. II.I.II.). In Fig. 11, several different magnetic field renderings of an FFL MPI scanner [53] are shown.

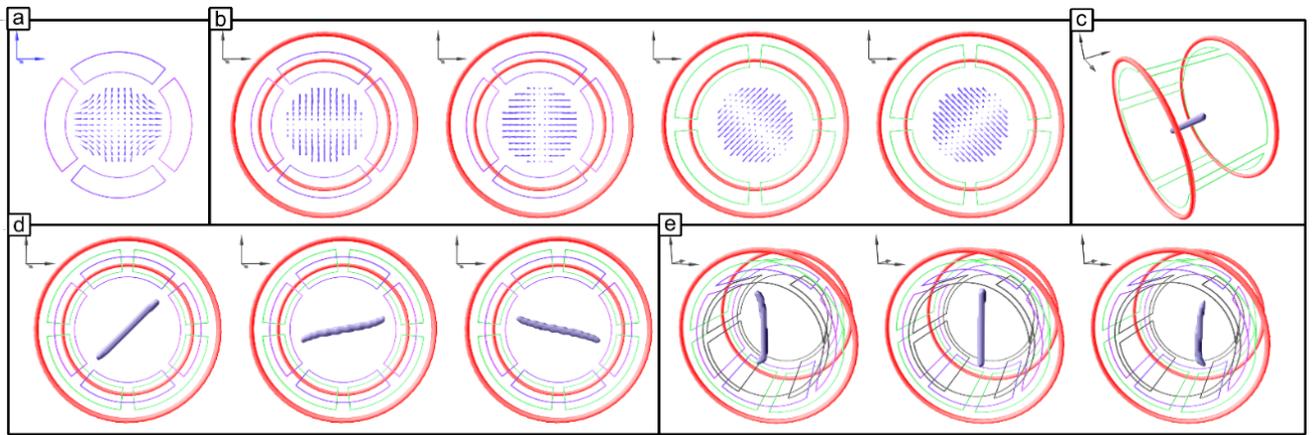

*Fig. 11: Renderings of magnetic field configurations for an FFL MPI scanner [53]. (a) starting with four saddle coils creating an FFL along the symmetry axis over magnetic field vector visualization of the FFL for different current settings (b) to surface renderings of the FFL (c) rotating about the symmetry axis (d) and displacing along the x-y slice (e).*

- *(Virtual) phantom generation*

    For investigation of imaging results or for comparison with realistic phantoms, it is useful to create virtual models of them. As described above (see Sec. II.I.VI.), an array of MNPs is calculated lying inside a given mesh model to obtain a filled 3D model with MPI generating signals, e.g., for testing realistic vessel structures within a virtual MPI scanner.

    In Fig. 12, an entire process of data extraction from 3D MRI or CT data sets over MPI simulation to measuring with realistic models is shown [54]. The desired structure has been extracted in a first step using a home-built framework providing an easy-to-use 3D GUI (Fig. 12 (1)+(2)) [55, 56]. The generated mesh mode can directly be used within the MFS software for virtual phantom generation (Fig. 12 (V1)+(V2)) and furthermore for creating real phantoms (Fig. 12 (R1)..(R4)), for example through 3D printing, coating, and washing [57]. Finally, both reconstructions can be compared (Fig. 12 (3)).

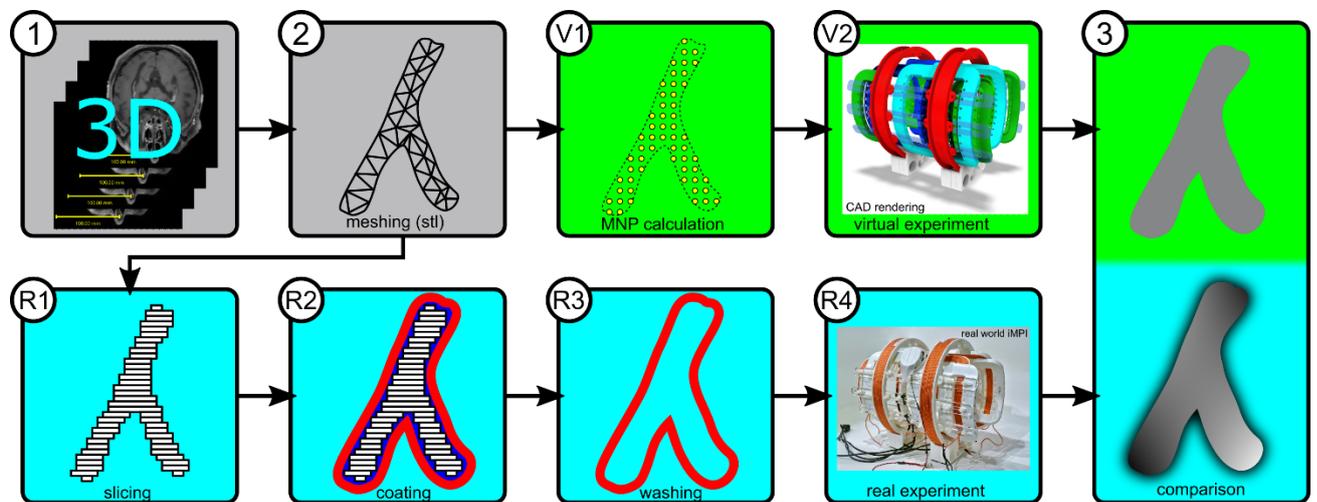

*Fig. 12: Sketch of the processing steps for generation of virtual&real phantoms for comparison: starting with the data extraction of, e.g., desired 3D vessel structures resulting in a 3D mesh model, the stl-file can be directly imported in MFS software for further virtual signal simulation (V1&V2). By 3D printing the stl-file followed by additional processing steps, a realistic phantom can be measured in a real MPI scanner (R1..R4).*

**II.II. Reconstruction Framework – RiFe**

The reconstruction framework (RiFe) is a software package, which provides a fast data processing of data sets either coming directly from an analog-digital converter like an MPI scanner or a simulation tool like MFS (see Sec. II.I.) via MMF or a proprietary data format (see Sec. VI.S2). It is optimized for latency times belows 100 ms between starting the sequence for collecting the data and visualizing them (see Fig. 13) [7]. This is mandatory for medical applications under real-time conditions, such as PTA or stenting [45, 46], to ensures fast visual feedback for the medical operator.

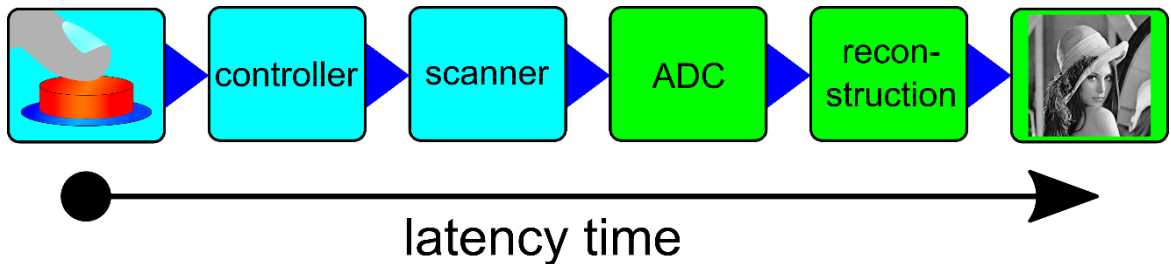

***Fig. 13:*** *Sketch of the important steps starting from pressing-the-button to the final reconstructed image. The latency time is the sum of all time lags of the corresponding steps [8].*

RiFe offers fast data manipulation as required, i.e., for receive-chain correction (RCC) of real MPI scanners, or frequency peak picking and filtering as preparation step for multiple reconstruction methods, such as image-based (re-gridding, direct deconvolution, image-based system matrix) and Fourier-based peak picking system matrix reconstruction [23, 25, 28, 29].

Different algorithms are available for the reconstruction process (see Fig. 14): Wiener filter, Kaczmarz algorithm, singular-value decomposition (SVD), etc., all optimized for near real-time calculation. The system matrix data can be generated directly within the MFS environment (see Sec. II.I.) for multiple scanner and reconstruction types.

As indicated in Fig. 14, the entire reconstruction process consists of several parts:

- ***Raw data***
  The raw data are generated either within the simulation software, e.g., MFS (see Sec. II.I.), or within a real MPI scanner and provided to the RiFe software.
- ***Data correction***
  From dedicated signal filtering in time and Fourier domain to data correction with received correction data sets (receive chain correction – RCC) there are multiple implementations available optimized for real-time data processing.
- ***Gridding***
  Based on the trajectory, the real (or virtual) MPI scanner has acquired the data, a raw-image or raw-volume can be generated (x-space).
- ***Reconstruction***
  Using deconvolution algorithms, 2D raw images can be reconstructed using, e.g., Wiener filter, or dedicated system matrices (model based or measured) for iterative image reconstruction.

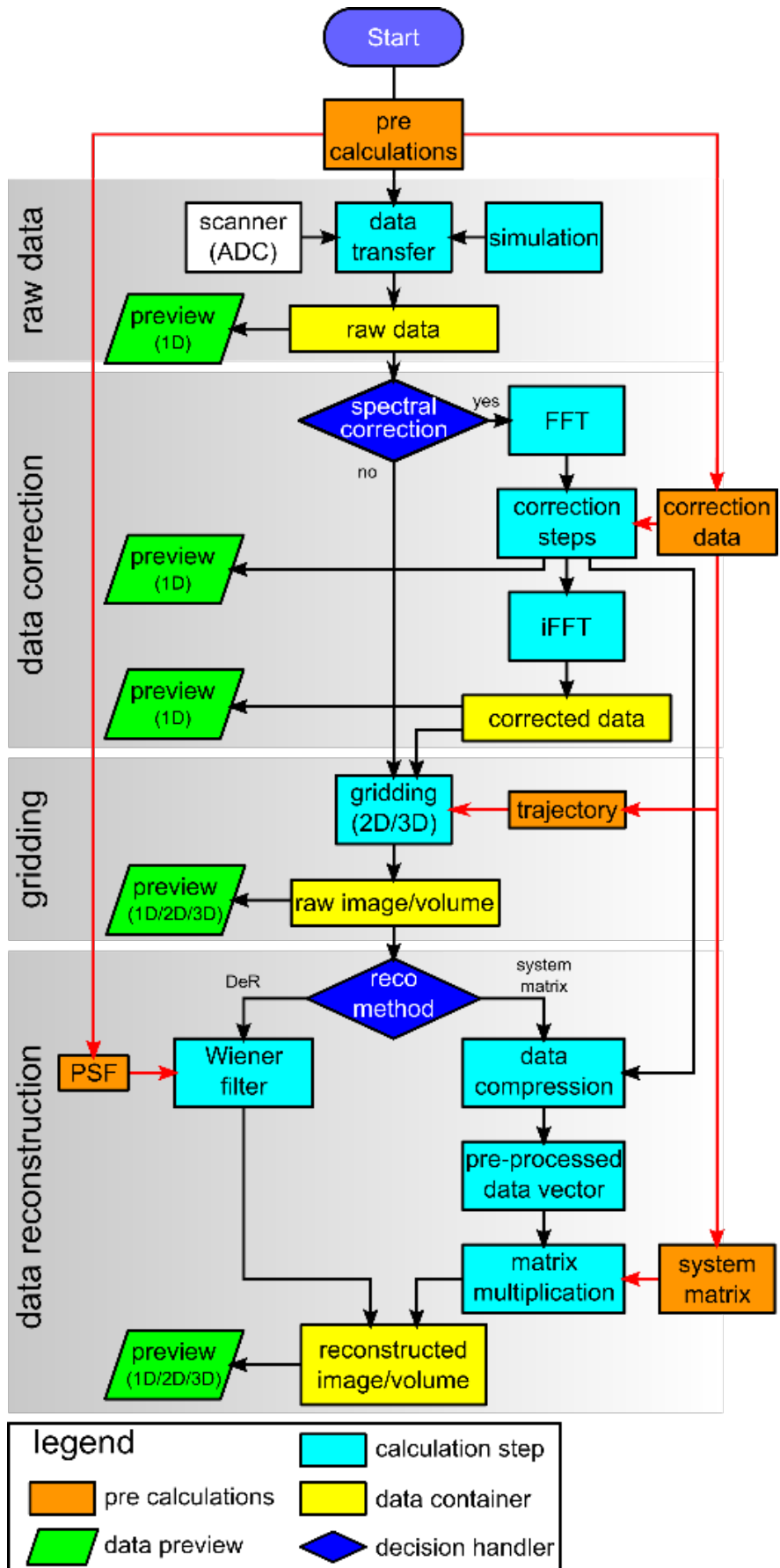

***Fig. 14:*** *Flow chart indicating the entire reconstruction process, which is separated into several sections: pre-calculation step, raw data step, correction step, gridding step, reconstruction step, and visualization step [7].*

A built-in graphic tool allows the direct visualization of time-signal, Fourier transformed spectra, Hysteresis, peak-picking and tracking, 2D raw-images and 2D reconstructed images with data transfer rates up to 67 frames per second depending on the data size and reconstruction settings [7]. For 3D visualization, data can be transferred via MMF data transfer to the 3D visualization tool (Sec. II.III.).

## II.III. 3D Visualization Tool – VT

Since the visualization of 2D data can be performed directly within the RiFe software, the visualization 3D data requires a more sophisticated rendering. The 3D visualization tool is a fast OpenGL volume renderer for direct visualization of 3D data sets using shader technology [58]. As data pipelines, the open-source data format NiFTI [59] is used, or a direct data transfer via MMF from the reconstruction framework (see Sec. II.II.) is provided.

For the latter data interface, additional information has been provided to map and visualize the 3D data correctly. Since the data stream is sent linearly, as a 1D array consisting of double values representing the brightness level, for each value the correct position in space must be provided, e.g., as an additional array of vectors. As indicated in Fig. 15, a 3D data set is reconstructed within the RiFe software (see Sec. II.I.) resulting in a 2D image. In this case, the 3D data are projected onto the 2D space, row-wise, where each row can be seen as slice [24]. With an appropriate coordinate look-up-table the 2D data can be mapped in the 3D space [24, 60].

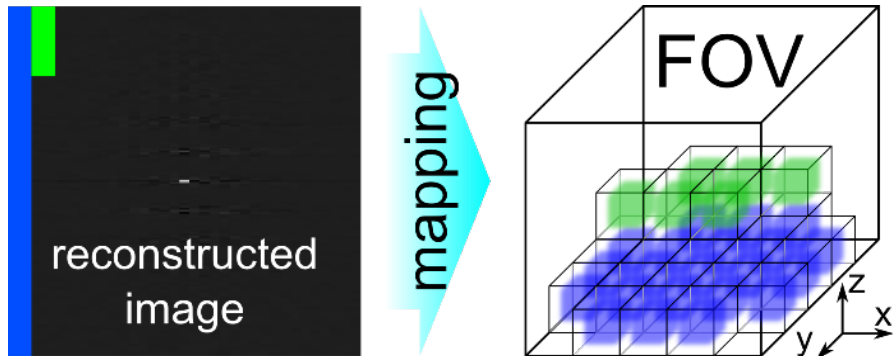

*Fig. 15:* Example of a 3D data set reconstructed and projected onto a 2D image. As indicated, each row of the 2D image represents a slice in the 3D volume. Using an appropriate coordinate look-up-table the 2D data can be mapped in a 3D space [24].

## III. Results

### III.I. 3D GUI MFS

Initially, a first overview of the features of the 3D GUI of the MFS software is given in Fig. 16, which shows several screenshots of the tutorial video (supplementary video 1: 3D-GUI). The fast and intuitive handling allows the creation and manipulation of different objects directly in 3D with fast feedback providing an easy-to-use environment for high flexibility and creativity.

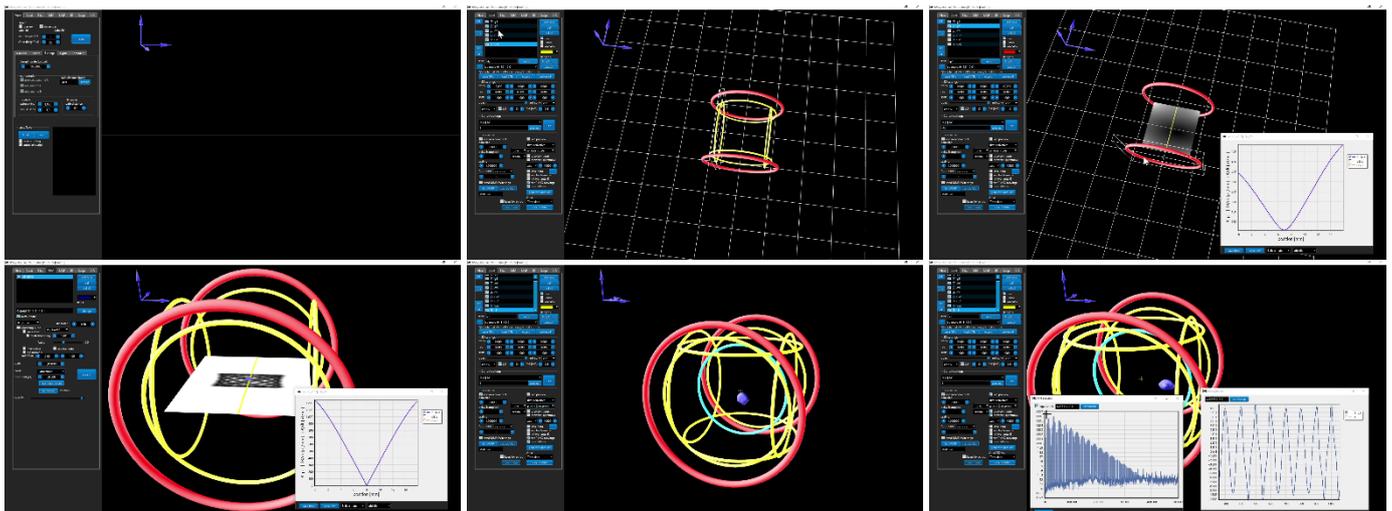

*Fig. 16:* Selected screenshots of a tutorial video for demonstrating the easy-to-use 3D GUI of the MFS software. Top: On the left, the starting window of the software is shown. Middle: after adding some solenoids, two (red) in Maxwell configuration creating a strong magnetic field gradient in their center, indicated as profile plot of the magnetic field (right), and additional two solenoid pairs in Helmholtz configuration for moving the resulting field free point (FFP) on a Lissajous trajectory through the FOV. Bottom left: the slice-container shows the stored trajectory. Bottom middle: an additional solenoid coil pair allows the displacement of the FFP along a 3D trajectory (FFP is rendered as blue 3D surface). Bottom right: with a receive coil (cyan), the signal can be rapidly calculated and showed as time-signal as well as Fourier transformed signal indicating the higher harmonics.

## III.II. Benchmark

The processes for calculating the magnetic field information have been optimized for a fast and intuitive user-experience with high response. For testing and quality demonstration, a benchmark has been implemented to test the performance of the MFS software on different platforms. The test consists of 12 solenoid conductors with 64 elements arranged as a dodecahedron each drive with a sinusoidal current. Using a single solenoid receive coil with 64 elements, a data set with 10k data points is generated of the 5569 MNP-container elements following the Langevin magnetization model (SPM). In Fig. 17, a screenshot of the test environment is shown, visualizing the magnetization at different time points. The results in Tab. 1 show the required calculation time on different machines.

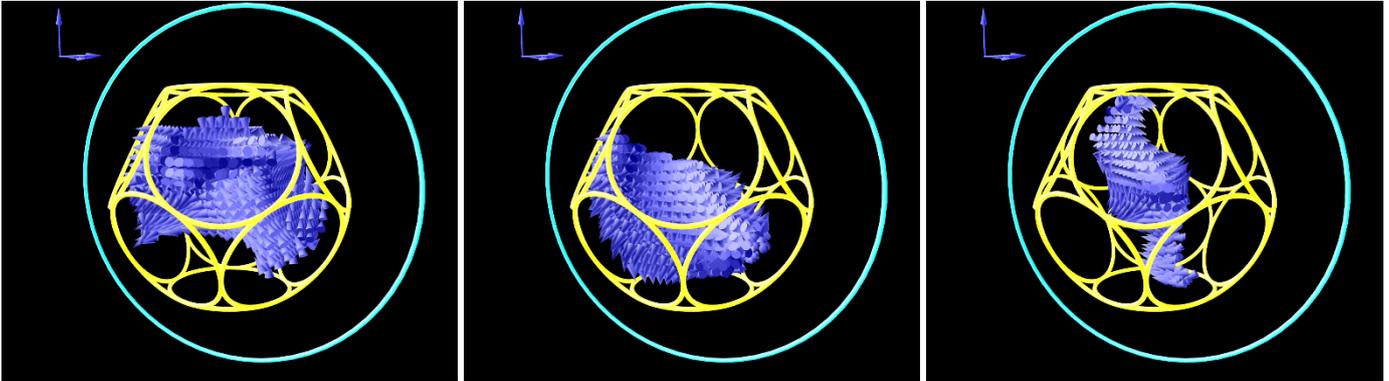

*Fig. 17: Screenshot of the test environment consisting of 12 solenoids and 5569 MNP-container visualized at different time points. The areas with low magnetic field are shown as blue arrows.*

*Tab. 1: Results of the MFS benchmark on different machines for single-threaded (st) and multi-threaded (mt) calculation. The calculation times have been averaged over 5 passages (software version: V.5.22.79[5]).*

| CPU | st – calc. time [ms] (*dl*=1k) | mt – Calc. time [ms] (*dl*=1k) | mt – Calc. time [ms] (*dl*=10k) |
|---|---|---|---|
| Intel i7-8650U (4 cores/8 threads) | 12533 | 715 | 7275 |
| Intel i7-1065G7 (4 cores/8 threads) | 13299 | 649 | 6442 |
| Intel i5-9500 (6 cores) | 12944 | 597 | 5875 |
| 2×Intel XEON E5-2670 (8 cores/16 threads) | 18588 | 2003 | 19195 |

## III.III. MFS projects

With the proposed software framework several different projects have been realized in the past, which are shortly introduced in the following. Starting with projects for educational purpose, e.g., to introduce different encoding schemes in MPI, additional tutorials are available to present the features of this software, especially the entire working process of the MFS+RiFe+VT framework. Finally, in the last section, a selection of scientific projects is shown, which are published in peer-reviewed journals.

- *Educational projects*

    One major goal of this framework was the visualization of magnetic fields for educational purpose. Especially the idea of MPI can be demonstrated in more detail as the selected projects demonstrated in the following.

    o *First MPI scanner á la Gleich&Weizenecker (G/W MPI)*

        In Fig. 18, an early simulation for a 2D MPI scanner á la Gleich and Weizenecker [1] is shown with an implemented 2D Lissajous trajectory scanning the FOV.

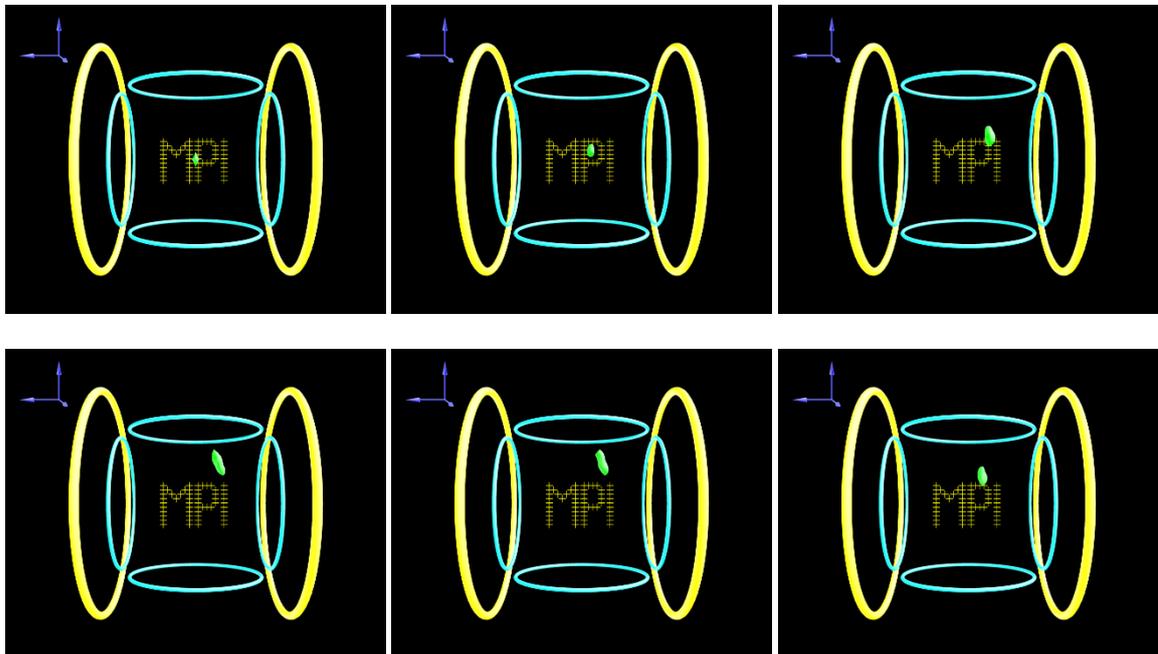

***Fig. 18:*** *Renderings of the first time steps showing the FFP moving along a 2D Lissajous trajectory through the FOV covering the 'MPI' logo. This concept was the starting point of MPI in 2005 [1].*

- **First FFL MPI design**

  In 2008, Weizenecker et al. proposed a novel kind of encoding scheme for Magnetic Particle Imaging. Instead of using an FFP, a field free line (FFL) can be used to acquire projection information of the sample under dedicated angles [61]. But the suggested approach was technically too challenging for realization. Erbe et al. could demonstrate an enhancement of the coil configuration, which allows a realistic implementation [53]. In Fig. 19, a simulation of the initially proposed FFL scanner is shown, the enhanced system can be seen in Fig. 11.

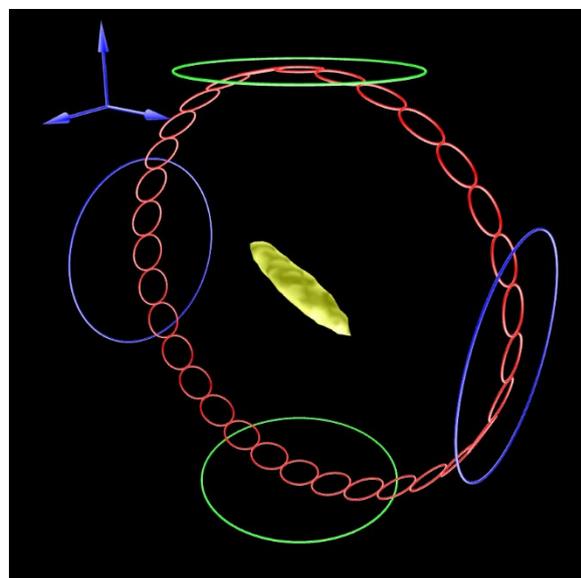

***Fig. 19:*** *Simulation of the first approach of FFL MPI scanner proposed by Weizenecker et al. [61]. The yellow surface rendering represents the FFL. The concept was too complicated to realize and would require multiple times more power than an FFP scanner with an effective field of view of similar size.*

- **Mech Halbach MPI scanner**

  The introduction of Halbach rings into MPI opened a door to new and promising scanner designs providing the advantage of generating strong magnetic field gradients without electrical power and thus heat dissipation [32, 62-65]. By constructing scanners composed of multiple different types of Halbach rings with individual axial rotations, different encoding schemes in form of FFPs or FFLs can be generated and moved through the FOV. A novel concept uses additional degrees of freedom, i.e., the poloidal and toroidal rotation of the Halbach rings to provide a higher flexibility. In Fig. 20, the renderings of the concept of such mechanically driven MPI scanner is shown.

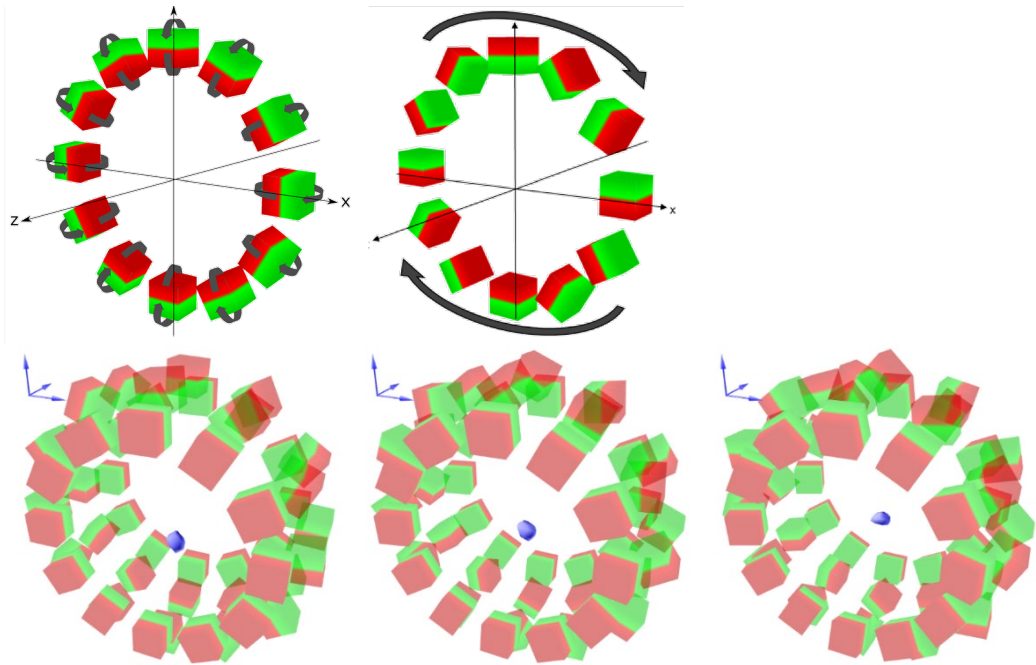

*Fig. 20: Renderings of the basic idea behind mechanically driven MPI scanner using Halbach rings. Top: poloidal and toroidal rotation of each magnet provides more degrees of freedom. Bottom: the combination of poloidal and toroidal rotated Halbach rings creates an FFP traveling along a 3D trajectory through the FOV.*

- o *Additional tutorials*

  For training purposes of new users of the entire framework, multiple tutorial videos have been created (see Sec. VI.S3. Tutorial videos). A special tutorial video demonstrates the interaction between all proposed software packages (MFS+RiFe+VT) starting with the simulation of a novel type of TWMPI scanner, followed by the generation of an appropriate 3D system matrix using peak-picking method (MFS), followed by data processing, reconstruction (RiFe), and visualization (VT). In Fig. 21, several screenshots of the key moments of the tutorial video are shown (see supplementary video 3: real-time visualization).

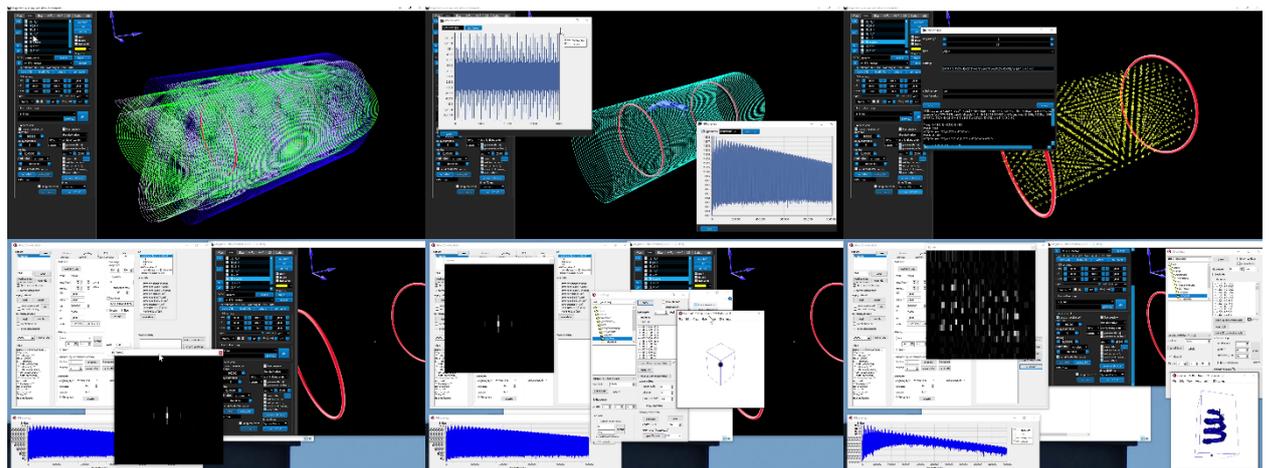

*Fig. 21: Several screenshots of the tutorial video demonstrating a fully virtual MPI FFL [66] scanner simulation, from data generation, system matrix creation, reconstruction and visualization with the proposed framework.*

- **Scientific projects**

  Using a powerful simulation software, such as MFS, several interesting approaches and novel ideas can be investigated, visualized, and understood before transfer into the real world.

  - o *Traveling Wave Magnetic Particle Imaging – TWMPI*

    The first scientific simulation project using MFS software was the Traveling Wave MPI system, which is an alternative scanner approach using a dynamic linear gradient array (dLGA) [67] for the generation and

movement of the FFP [25]. In first studies, the scanner geometry and size could be investigated before real implementation (see Fig. 21).

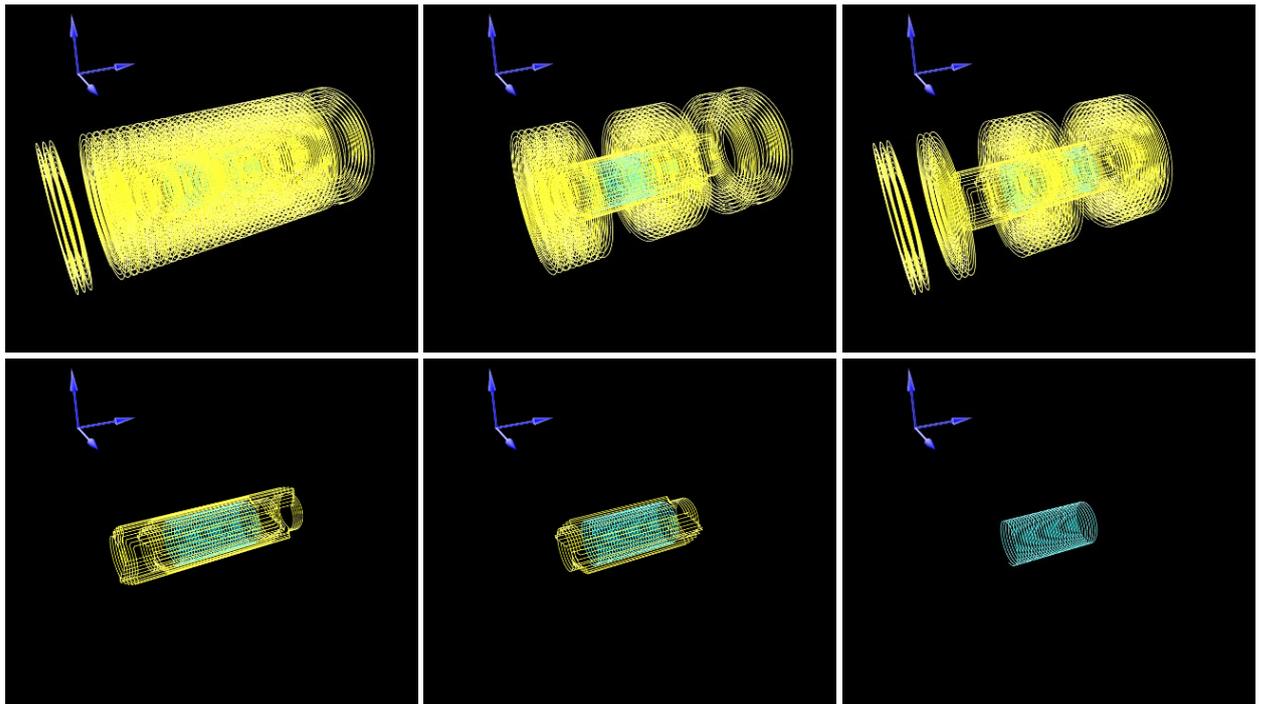

**Fig. 21:** *Renderings of the different coil parts the TWMPI scanner consists of: the dLGA consisting of 20 solenoids combined in two channels (top), two perpendicular saddle-coil pairs and a receive coil (bottom).*

o ***MPI meets CT***

This project combined for the first time an MPI scanner with a CT system. For that a novel design has been used based on Halbach arrays for the generation of a static field free line (FFL) [63]. With additional solenoids, the FFL can be moved within the imaging slice (See Fig. 22). In combination with a mechanical rotation of the sample, a full 2D reconstructed image can be provided simultaneously to the CT experiments [62].

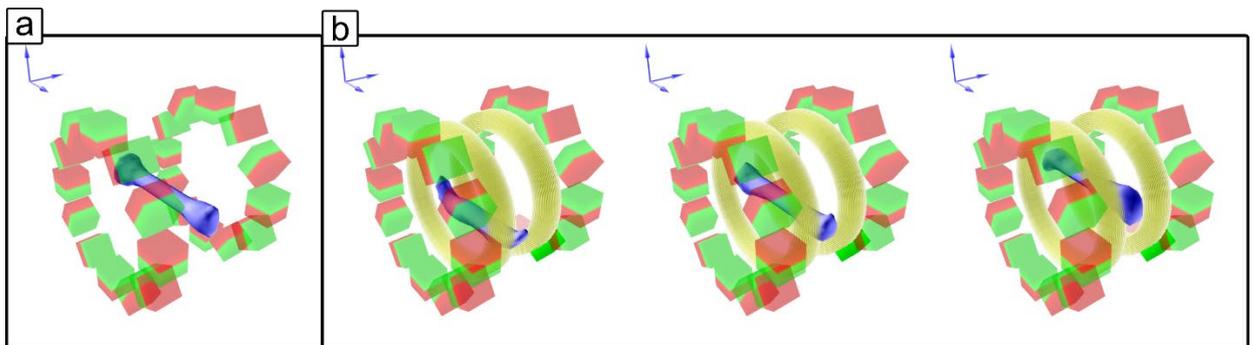

**Fig. 22:** *renderings of the basic concept behind the FFL MPI scanner based on Halbach rings. (a) using two Halbach rings, a static FFL can be generated in their center. (b) additional solenoids can move the FFL rapidly through a 2D slice for imaging.*

o ***Parallel MPI – pMPI***

The Traveling Wave Magnetic Particle Imaging scanner [25] allows data acquisition with multiple encoding schemes (here multiple FFPs) simultaneously using multiple receive coils. In initial simulation studies emulating the data acquisition, reconstruction process and visualization, the pMPI approach could be validated in simulations (see Fig. 23) and real experiments [68].

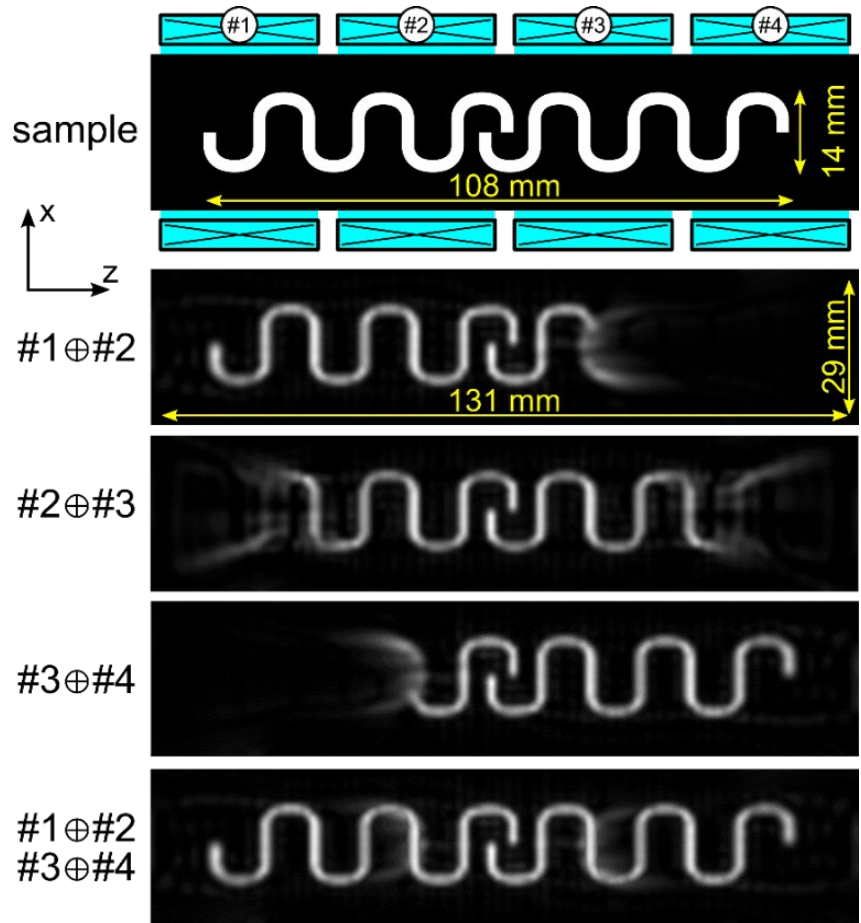

***Fig. 23:*** *Result of the simulation study of the parallel MPI approach using two field free points (FFPs) for data acquisition simultaneously [68].*

- **Zoom MPI**

    The dynamic linear gradient array (dLGA [67]) of a TWMPI scanner allows an adjustable gradient strength yielding a 'zoom' effect enhancing the spatial resolution within a specific region of the FOV [69]. All simulation studies have been performed with the mentioned MFS software (see Fig. 24).

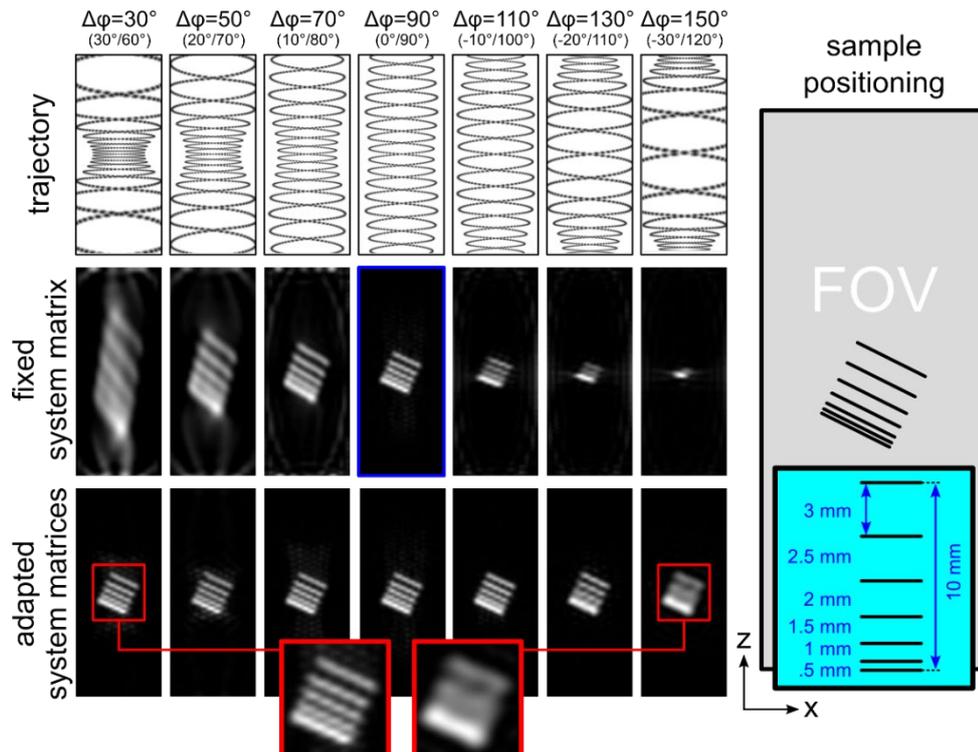

***Fig. 24:*** *Simulation of the 'zoom' effect provided by TWMPI scanners through adjusting the phase between the main gradient system [69].*

- *iMPI*
  
  The goal of the iMPI (interventional Magnetic Particle Imaging) project is the implementation of a first human-sized MPI scanner for a human leg dedicated for interventional treatment, such as real-time PTA or stenting [70, 71]. In prior simulation studies, the performance of such a novel kind of scanner has been investigated [72].

  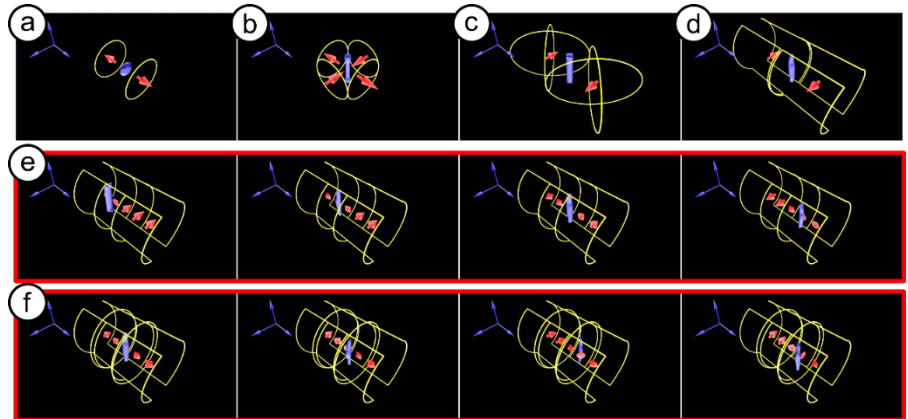

  *Fig. 25: Evolution from a FFP MPI design (a) to a novel kind of FFL scanner providing a compact system for human-sized MPI-guided treatment (f) [72].*

- *COMPASS*
  
  The idea behind COMPASS (Critical Offset Magnetic PArticle SpectroScopy) is the measurement of slightly changes in the hydrodynamic diameter of functionalized MNPs, caused by binding for example SARS-CoV-2 antibodies, providing an MNP-based rapid test with sensitivities comparable to ELISA. Performing simulation studies with MFS software, the basic effect behind this method was investigated in more detail (see Fig. 26) [73].

  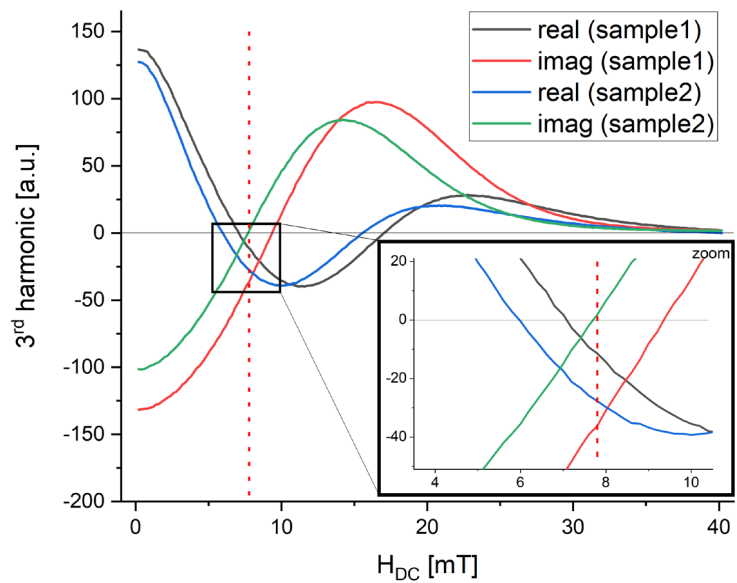

  *Fig. 26: Result of an offset magnetic field sweep simulation of two samples with slightly different $\zeta$-parameter using Langevin equation. Both data sets show differences in the position of the crossing points of real and imaginary curves (critical point on the $3^{rd}$ harmonic), which can be used for highly sensitive measurement of phase differences between both samples.*

- *3D system function visualization*
  
  For understanding the reconstruction process in MPI, the specific system functions for MPI scanner can be investigated [74]. Using the proposed framework (MFS+RiFe+VT), the system functions for a specific sequence for TWMPI scanner can be investigated and visualized (see Fig. 27) [24].

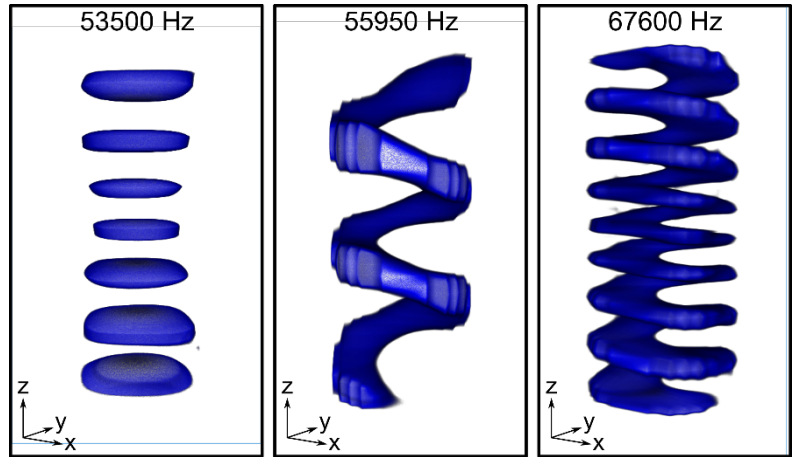
*Fig. 27: 3D volume renderings of selected system functions of a specific 3D sequence for TWMPI scanners [24].*

## IV. Discussion

Fast calculation processes are of high interest in many fields, especially for real-time applications such as visualization with low latency. The proposed framework consists of three software packages, each optimized for a specific task: MFS for simulation of magnetic fields and non-linear magnetization response, RiFe for signal processing and image reconstruction, and VT for volume rendering of 3D data. In combination, the framework is optimized for real-time visualization. An example is RiFe & VT in combination with a real MPI scanner [24, 45, 46, 72].

There are many possibilities for optimizing the source-code for better performance as discussed in [7] for RiFe software, e.g., using optimized libraries for matrix calculation, Fourier transformation or outsourcing on GPUs (graphic processing unit). For VT software, regular updating the 3D hardware and driver can boost the performance of the visualization dramatically, but in the end, the bottleneck here are the previous calculation processes (MFS & RiFe).

Despite several optimizations, the generation of data sets within MFS software can take a while and especially when generating system matrices with a high number of data points, the waiting times can be quite long (c.f. tutorial video – Fig. 21). To speed up the calculation process, an outsourcing of specific calculation steps on the GPU is possible, which provides a massive parallelization capability [75].

Further optimizations have to be performed to use the provided software packages on computers with more than 100 cores, such as memory optimization to reduce CPU-RAM transfer rates, clustering calculation packages for NUMA cores, etc.

For a higher flexibility and user experience, a server/client system can be implemented to entangle the calculation process from the visualization process. Since these processes did not work on the same machine anymore, an easy-to-use environment for fast reconstruction visualization without the need of powerful computers are possible.

To extend the possibilities of the simulation framework (MFS), several magnetization models can be added for a more realistic signal calculation, e.g., the mentioned Langevin equation is just valid for Brownian relaxation and does not cover Néel relaxation of MNP ensembles.

Furthermore, realistic descriptions of the used component, here the conductor-container, can be useful for a more realistic simulation of magnetic field generators. Parameters such as self-inductance, inductivity or blind capacities, which yields eigen-frequencies in the simulated systems, are useful to calculate transfer functions or receive chains with a higher precision. This provides a better understanding of signal generation and data processing in MPI.

Finally, fully automated and intelligent algorithms can be used for optimization processes of magnetic fields, magnetic field gradients, conductor distributions, or signal processing to fit a specific question. The latest achievements in AI-based methods, such as machine learning, evolutionary algorithms, etc. show promising results and could help to support the user of such software packages.

## V. Conclusion

The presented framework covers the entire procedure from simulation of a 3D MPI scanner (design, hardware, magnetic field simulation) using MFS software over the reconstruction software (RiFe) offering real-time reconstruction and visualization to 3D visualization (VT software) using volume rendering. The easy-to-use interface, the short calculation times, and the modular processing chain, allows fast prototyping of novel scanner approaches as well as the comparison with real data of implemented MPI systems.

## VI. Supplementary material

### S1. 3D affine transformation matrix
Any combination of translation $T$, rotation $R$, and scaling $S$ of a 3D object can be combined within a single $4 \times 4$ affine transformation matrix $\widehat{M}$.

$$\widehat{M} = R_x(\vartheta) \cdot R_x(\vartheta) \cdot R_x(\vartheta) \cdot S \cdot T$$

$$R_x(\vartheta) = \begin{pmatrix} 1 & 0 & 0 & 0 \\ 0 & \cos(\vartheta) & \sin(\vartheta) & 0 \\ 0 & -\sin(\vartheta) & \cos(\vartheta) & 0 \\ 0 & 0 & 0 & 1 \end{pmatrix}$$

$$R_y(\vartheta) = \begin{pmatrix} \cos(\vartheta) & 0 & -\sin(\vartheta) & 0 \\ 0 & 1 & 0 & 0 \\ \sin(\vartheta) & 0 & \cos(\vartheta) & 0 \\ 0 & 0 & 0 & 1 \end{pmatrix}$$

$$R_z(\vartheta) = \begin{pmatrix} \cos(\vartheta) & -\sin(\vartheta) & 0 & 0 \\ \sin(\vartheta) & \cos(\vartheta) & 0 & 0 \\ 0 & 0 & 1 & 0 \\ 0 & 0 & 0 & 1 \end{pmatrix} \quad (7)$$

$$S = \begin{pmatrix} s_x & 0 & 0 & 0 \\ 0 & s_y & 0 & 0 \\ 0 & 0 & s_z & 0 \\ 0 & 0 & 0 & 1 \end{pmatrix}$$

$$T = \begin{pmatrix} 1 & 0 & 0 & t_x \\ 0 & 1 & 0 & t_y \\ 0 & 0 & 1 & t_z \\ 0 & 0 & 0 & 1 \end{pmatrix}$$

### S2. Used data formats
There are multiple data formats used within the framework:

- **Format for MFS project files (*.mfs5)**
  Proprietary file format for MFS software for storing the settings for an MFS project (readable ASCII text-file).
- **Format for MFS-script files (*.mf5s)**
  Proprietary file format for MFS scripting command lists (readable ASCII text-file).
- **Format for connected conductor files (*.cnl)**
  Proprietary file format for MFS software for storing connected conductor vector information (readable ASCII file).
- **Format for 3D vectors (*.3dv)**
  Proprietary file format for MFS software to store MNP array (binary file: 4 Byte header with number $N$ of entries, followed by $N \times 3 \times$ double (8 byte) data array for [x,y,z]).
- **Format for 3D mesh models (*.stl)**
  Open-source file format for 3D mesh models (standard triangle language).
- **Format for data sets (*.mpi & *.dbl)**
  Proprietary file format for storing data sets: the *.mpi file is a readable ASCII file consisting of the name of the data set, the samplingrate, and the name of the connected file (*.dbl), which consists of the data set information (binary file: no header with $N \times$ double (8 bytes) values; $N$ can be determined by the size of the *.dbl-file).
- **Format for system matrices (*.sm, *.2da, *.1da & *.txt)**
  Several proprietary file formats for storing a full system matrix (*.sm) or 2D and 1D arrays (*.2df and *.1df). The text-files (*.txt) consists of all required system matrix information, such as matrix sizes and type. The *.sm and *.2df file are binary files with header: *sizeX* (integer) and *sizeY* (integer) followed by

*sizeX*×*sizeY*×double (8 bytes) data array consisting of the values. The *.1df file stores a 1D array (binary file with header: *size* (integer) followed by *size*×double (8 bytes) data array.

- **Format for 3D data sets (*.nii)**
  This data format is an open-source data format and follows the NiFTi specifications [59].
- **Formats used for RiFe (*.2df, *.txt, *.png, *.bmp)**
  The *.2df file if a proprietary file format for storing 2D field information (binary file: header (32 bytes) with *sizeX* (integer – 4 bytes), *sizeY* (integer – 4 bytes), *maxValue* (double – 8 bytes), *minValue* (double – 8 bytes), *maxCoordX* (integer – 4 bytes), *maxCoordY* (integer – 4 bytes) followed by *sizeX*×*sizeY*×double (8 bytes) data array.
  The *.txt file consists of settings information for RiFe software.
  The *.png and *.bmp file formats are used as standard formats for pictures.

*S3. Tutorial videos*

Multiple tutorial videos area available for training purposes introducing the key features of the proposed framework. A complete list can be found in the latest version of the MFS software.

- 1. Tutorial video 1: first steps in MFS ([link](link))
- 2. Tutorial video 2: transmit&receive ([link](link))
- 3. Tutorial video 3: conductors ([link](link))
- 4. Tutorial video 4: coil-paint ([link](link))
- 5. Tutorial video 5: conductor wizard ([link](link))
- 6. Tutorial video 6: combine conductors ([link](link))
- 7. Tutorial video 7: MPI scanner (FFP) ([link](link))
- 8. Tutorial video 8: MPI scanner (FFL) ([link](link))
- 9. Tutorial video 9: TWMPI scanner ([link](link))
- 10. Tutorial video 10: 3D mesh to MNP distribution ([link](link))
- 11. Tutorial video 11: MFS5 script ([link](link))
- 12. Tutorial video 12: Rodin-coil ([link](link))
- 13. Tutorial video 13: Halbach MPI ([link](link))
- 14. Tutorial video 14: Halbach MPI (MPI meets CT) ([link](link))
- 15. Tutorial video 15: real-time visualization (german) ([link](link))
- 16. Tutorial video 16: trajectory trend ([link](link))
- **Supplementary video 1**: Introduction to the 3D-GUI
- **Supplementary video 2**: Introduction to system matrices
- **Supplementary video 3**: Real-time visualization using MFS+RiFe+VT

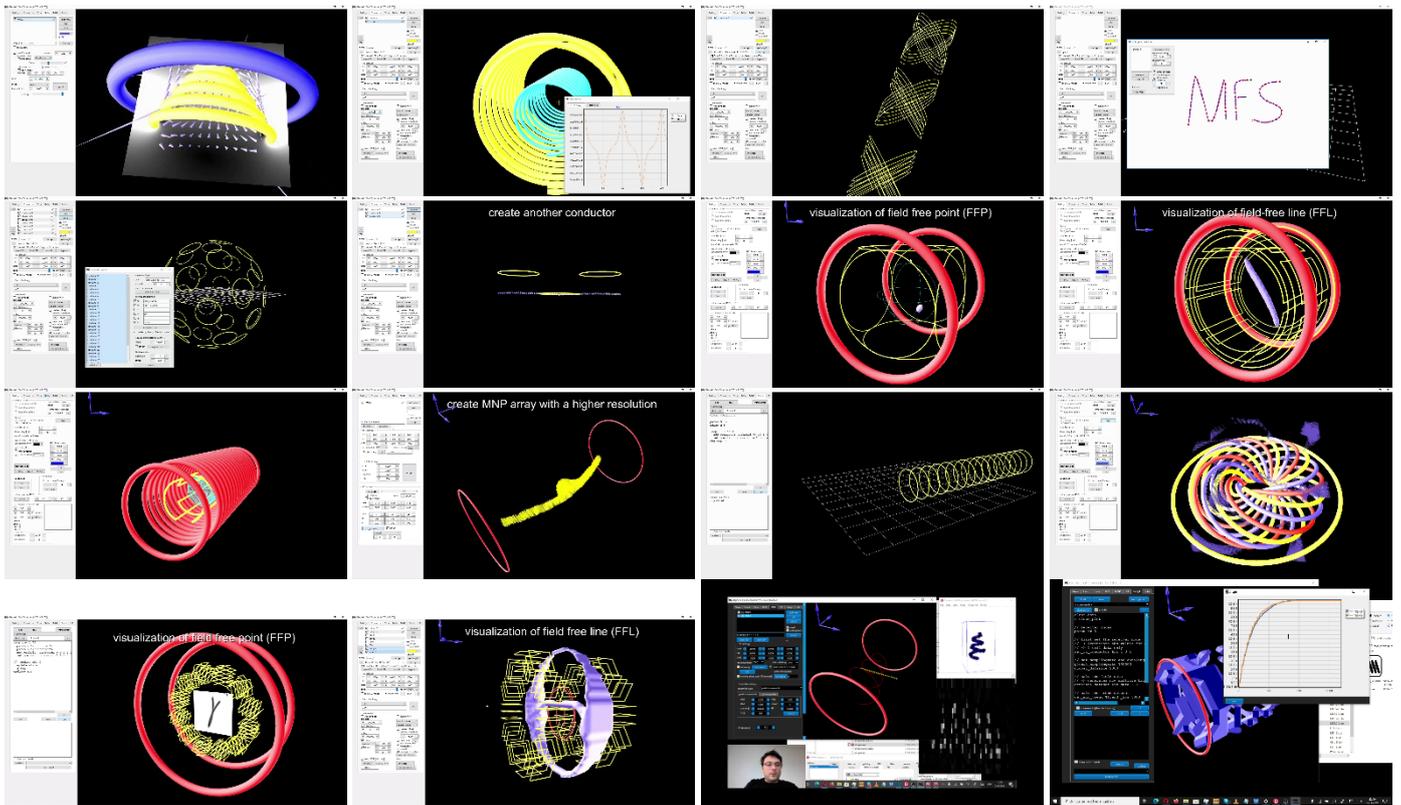

*Fig. S3:* Several screenshots of the tutorial videos available so far for MFS software and combination of MFS+RiFe+VT (tutorial videos 1..16).

**AUTHOR'S STATEMENT**
Research funding: This work was partially funded by the DFG (BE 5293/1-1, BE 5293/1-2, VO 2288/1-1, VO 2288/2-1, VO 2288/3-1) and the IDEA project of the 7[th] framework programme of the European Union (project reference 279288). Conflict of interest: Authors state no conflict of interest.